\documentclass[fleqn,usenatbib]{mn2e}
\pdfoutput=1
\usepackage{mathptmx}
\usepackage[T1]{fontenc}
\usepackage{ae,aecompl,pdflscape}

\usepackage{graphicx}	% Including figure files
\usepackage{amsmath}	% Adv\usepackage{longtable}anced maths commands
\usepackage{amssymb}	% Extra maths symbols
\usepackage{float}
\usepackage{epstopdf}
\usepackage{subfigure}
\usepackage{ulem}
\usepackage{array}
\usepackage{booktabs}
\usepackage{supertabular}
\usepackage{multicol} 
\title[AGN winds and H$_2$ emission]{Active Galactic Nuclei Winds as the Origin of the H$_2$ Emission Excess in Nearby Galaxies}

\author[Riffel, Zakamska \& Riffel]{Rogemar A. Riffel,$^{1,2}$\thanks{E-mail: rogemar@ufsm.br (RAR)}
Nadia L. Zakamska,$^{1}$
Rog\'erio Riffel$^{3}$
\\
$^{1}$Department of Physics \& Astronomy, Johns Hopkins University, Bloomberg Center, 3400 N. Charles St, Baltimore, MD 21218, USA\\
$^{2}$Universidade Federal de Santa Maria, CCNE, Departamento de F\'\i sica,  
 97105-900, Santa Maria, RS, Brazil\\
$^{3}$Universidade Federal do Rio Grande do Sul, IF, CP 15051, Porto Alegre 91501-970, RS, Brazil
}

\date{Accepted XXX. Received YYY; in original form ZZZ}

\pubyear{2019}

\begin{document}
\label{firstpage}
\pagerange{\pageref{firstpage}--\pageref{lastpage}}
\maketitle

% Abstract of the paper
\begin{abstract}
In most galaxies, the fluxes of rotational H$_2$ lines strongly correlate with star formation diagnostics (such as polycyclic aromatic hydrocarbons, PAH), suggesting that H$_2$ emission from warm molecular gas is a minor byproduct of star formation. 
We analyse the optical properties of a sample of 309 nearby galaxies derived from a parent sample of 2,015 objects observed with the {\it Spitzer Space Telescope}. We find a correlation between the [O\,{\sc i}]$\lambda$6300 emission-line flux and kinematics and the H$_2$S(3)\,9.665\,$\mu$m/PAH\,11.3\,\,$\mu$m. The [O\,{\sc i}]$\lambda$6300 kinematics in Active Galactic Nuclei (AGN)  can not be  explained only by gas motions due to the gravitational potential of their host galaxies, suggesting that AGN driven outflows are important to the observed kinematics. While H$_2$ excess also correlates with the fluxes and kinematics of ionized gas (probed by [O\,{\sc iii}]), the correlation with [O\,{\sc i}] is much stronger, suggesting that H$_2$ and [O\,{\sc i}] emission probe the same phase or tightly coupled phases of the wind.
We conclude that the excess of H$_2$ emission seen in AGN is produced by shocks due to AGN driven outflows and in the same clouds that produce the [O\,{\sc i}] emission. Our results provide an indirect detection of neutral and molecular winds and suggest a new way to select galaxies that likely host molecular outflows. Further ground- and space-based spatially resolved observations of different phases of the molecular gas (cold, warm and hot) are necessary to test our new selection method.
\end{abstract}

% Select between one and six entries from the list of approved keywords.
% Don't make up new ones.

%%%%%%%%%%%%%%%%%%%%%%%%%%%%%%%%%%%%%%%%%%%%%%%%%%

%%%%%%%%%%%%%%%%% BODY OF PAPER %%%%%%%%%%%%%%%%%%

\section{Introduction}\label{sec:intro}

\begin{table*}
\centering
\caption{The sub-samples.}
 \begin{tabular}{c c c l}
\hline
Name & \# of gal. in  & \# of matches  & Comments\\
     & \citet{lambrides}& in SDSS      & \\
     \hline
Sample Y & 485 & 115 & H$_2$S(3)\,9.665\,$\mu$m, and PAH\,11.3\,\,$\mu$m emission detected. \\
Sample N & 1503 & 193 & H$_2$S(3)\,9.665\,$\mu$m emission not detected. \\
Other & 27 & 1 & H$_2$S(3) detected, PAH not detected. \\
\hline
\end{tabular}
\label{tab:sample}
\end{table*}

Identifying and characterizing the processes that transform galaxies from star-forming to quiescent is a fundamental goal of extragalactic astronomy \citep{conselice14,hatfield17,liu18,kim18}. 
Some of the critical transformation mechanisms include galactic-scale feedback due to Active Galactic Nuclei (AGN) or star formation. This feedback is now thought to
 be extremely important for galaxies of all mass scales \citep{cattaneo09,alexander12,fabian12,harrison17}.  
Galactic ionized gas outflows driven by AGN  \citep{liu13,carniani15,fischer17}  or star formation \citep{arribas14,gallagher19}  have been mapped in the last decade, leading to major improvements in understanding galactic winds.  But what happens to the molecular gas is much less clear.  This component is of significant interest for understanding the impact of molecular outflows on star formation. 

Studying cold ($T\lesssim 100$ K), warm ($T\sim$a few hundred K) and hot  ($T\gtrsim1000$ K) phases of molecular gas is essential for understanding the origin and role of galactic molecular outflows. Studies of the inner kpc of nearby galaxies, using near-infrared integral field spectroscopy  assisted by adaptive optics systems on 8-10~m class telescopes,   have shown that hot molecular gas outflows are very scarce \citep{davies14,n5929,may18}. Usually, the hot H$_2$ emission arises from the circumnuclear rotationally supported gas disk, sometimes showing streaming motions towards the nucleus  \citep{n4051,m79,ms09,mazzalay14,durre19,schonell19}. 

Ultra-Luminous Infrared Galaxies (ULIRGs) seem to have an excess of hot molecular gas emission relative to that expected from their star formation rates \citep{zakamska10}, possibly due to shock-heating by supernova- or AGN-driven outflows \citep{hill14,imanishi18,imanishi19}. Indeed, recent near-infrared integral field spectroscopy reveals the presence of hot molecular gas outflows \citep{emonts17} in three of of four observed ULIRGs. Cold molecular gas outflows  are also commonly observed in powerful AGN \citep{feruglio10,fiore17}. In nearby ULIRGS and AGN host galaxies, cold molecular outflows have been detected using {\it Herschel } spectra in the far-infrared lines of OH \citep[e.g.][]{fischer10,veilleux13,gonzalez-alfonso14,gonzalez-alfonso17}  and spatially resolved with Atacama Large Millimeter Array (ALMA) observations of CO lines \citep[e.g.][]{combes13,gb14,morganti15,pereira-santaella18,rama19,alonso-herrero19,husemann19}. These studies reveal outflows with velocities ranging from few tens of km\,s$^{-1}$ to over 1\,000  km\,s$^{-1}$,  mass-outflow rates of up to 10$^3$ M$_\odot$\,yr$^{-1}$ and kinetic power as high as 10$^{44}$\, erg\,s$^{-1}$. Accelerating dense molecular gas to high enough velocities that they would escape the galaxy is extremely difficult, so modern theoretical work suggests that molecules may be formed within the outflow and may  display excitation characteristics of shock heating \citep{richings18a,richings18b}. 

Understanding the acceleration and the emission mechanisms of molecular outflows -- the critical ingredient in rapid star formation quenching -- remains a major unsolved problem in galaxy formation.  A recent study by \citet{lambrides} provide an important new tool for understanding warm (a few hundred K)  molecular gas emission in nearby galaxies. They analyse 2,015 mid-infrared (mid-IR) spectra of galaxies observed with the {\it Spitzer Space Telescope} and  provide fluxes of all emission features. Furthermore, they measure the excitation  temperature of the H$_2$S(5) and  H$_2$S(7) pure rotational transitions using stacked spectra of AGN-dominated and non-AGN dominated sources. They find that the H$_2$ fluxes are higher in AGN than in star-forming galaxies and the excitation temperature on AGN-dominated galaxies is $\sim$200\,K larger, indicating that the AGN plays an important role in the H$_2$ emission. However, due to the low resolution of the Spitzer spectra, there is currently no available information on the H$_2$ kinematics, necessary to investigate if the H$_2$ emission arises from shock-heated gas or if it is due to local heating of the gas by AGN or stellar radiation field.

 In this paper, we cross-match the sample used by \citet{lambrides} with the Sloan Digital Sky Survey \citep[SDSS; ][]{dr15}   spectroscopic database in order to investigate the origin of the warm molecular hydrogen emission.  
 This paper is organized as follows. Section~ \ref{sec:sample} presents and characterize the sample;  Sec.~3 describes the data compilation and measurements procedure. In  Sec.~4 we present our results, which are discussed in Sec.~5 and summarized in Sec.~ 6. 
 
 We use a $h=0.7, \Omega_m=0.3, \Omega_{\Lambda}=0.7$ cosmology. The wavelengths of the emission features in the infrared are given in vacuum. The wavelengths of emission lines in the optical are given in the air (e.g., [O\,{\sc iii}]$\lambda$5007\AA) following a long-standing tradition, even though the SDSS spectroscopic database uses vacuum wavelengths. To test whether two distributions are statistically consistent (i.e. whether they are drawn from the same underlying distribution), we use the Kolmogorov-Smirnov test and report the  probability of the null hypothesis $P_{\rm KS}$ that the two samples are consistent. A small value of $P_{\rm KS}$ implies a statistically significant difference in the distributions. To test whether two  parameters are correlated, we use the Pearson test to computed the $P_{\rm rank}$ value. A small value of $P_{\rm rank}$ implies in a statistically significant correlation between the parameters and we consider that there is  a correlation if  $P_{\rm rank}<0.05$.

\section{Data and Measurements} \label{sec:sample}
\subsection{The parent sample}

\citet{lambrides} analysed the  molecular gas properties of a sample of 2,015 galaxies ($0.002 < z < 3.0)$ using mid-infrared spectra obtained with the {\it Spitzer Space Telescope}. Their sample includes all objects observed as part of programs containing at least one of the following keywords in their abstracts: AGN, Radio Galaxy, QSO, Quasar, Starburst Galaxy, or ULIRG/LIRG, using the Infrared Spectrograph (IRS) in the low-resolution modules (SL and LL) and they have excluded spectra with detection levels $<3\sigma$. Each module is divided into two spectroscopic orders: SL1 (7.46\,$\mu$m $<\lambda<$ 14.29\,$\mu$m), SL2 (5.13\,$\mu$m $<\lambda<$ 14.29\,$\mu$m), LL1 (19.91\,$\mu$m $<\lambda<$ 39.90\,$\mu$m) and LL2 (13.90\,$\mu$m $<\lambda<$ 21.27\,$\mu$m).  

In this work, we cross-match the sample used by \citet{lambrides} with the SDSS Spectroscopic Database \citep{gunn06,blanton17}. The SDSS spectra cover the range 3600 -- 10300\,\AA\ at a resolving power $R\sim$2000 and are part of the fifteenth Data Release (DR15) of the SDSS project \citep{dr15}.  We use the SDSS Query/CasJobs platform to search for optical spectra of each object of the Lambrides sample. We include only objects photometrically identified as ``Galaxy" and located closer than 0\farcm1 from the Spitzer coordinates.  We find that 309 galaxies from the sample of \citet{lambrides} have spectra available in the DR15 of SDSS. 

We divide the parent Spitzer sample into two sub-samples: 
\begin{itemize}
    \item (i) {\it The sample Y} contains all objects with the H$_2$S(3)\,9.665\,$\mu$m and PAH\,11.3\,\,$\mu$m emission lines detected. These are the most commonly detected H$_2$ and star formation diagnostics in the Spitzer dataset.
%    These specific lines are chosen because they were used by \citet{lambrides} to characterize the H$_2$ emission, by defined the molecular hydrogen excess (H$_2$S(3)/PAH\,11.3\,\,$\mu$m line ratio). 
This sample contains 485 galaxies, with 115 matching the SDSS database. 

 \item (ii) {\it The sample N} includes 1503 galaxies for which the H$_2$S(3)\,9.665\,$\mu$m line was not detected in the Spitzer spectra. Spectra for 193 galaxies are available in the SDSS DR15. 
  \end{itemize}
  
\begin{figure*}
  \centering
\includegraphics[width=0.8\textwidth]{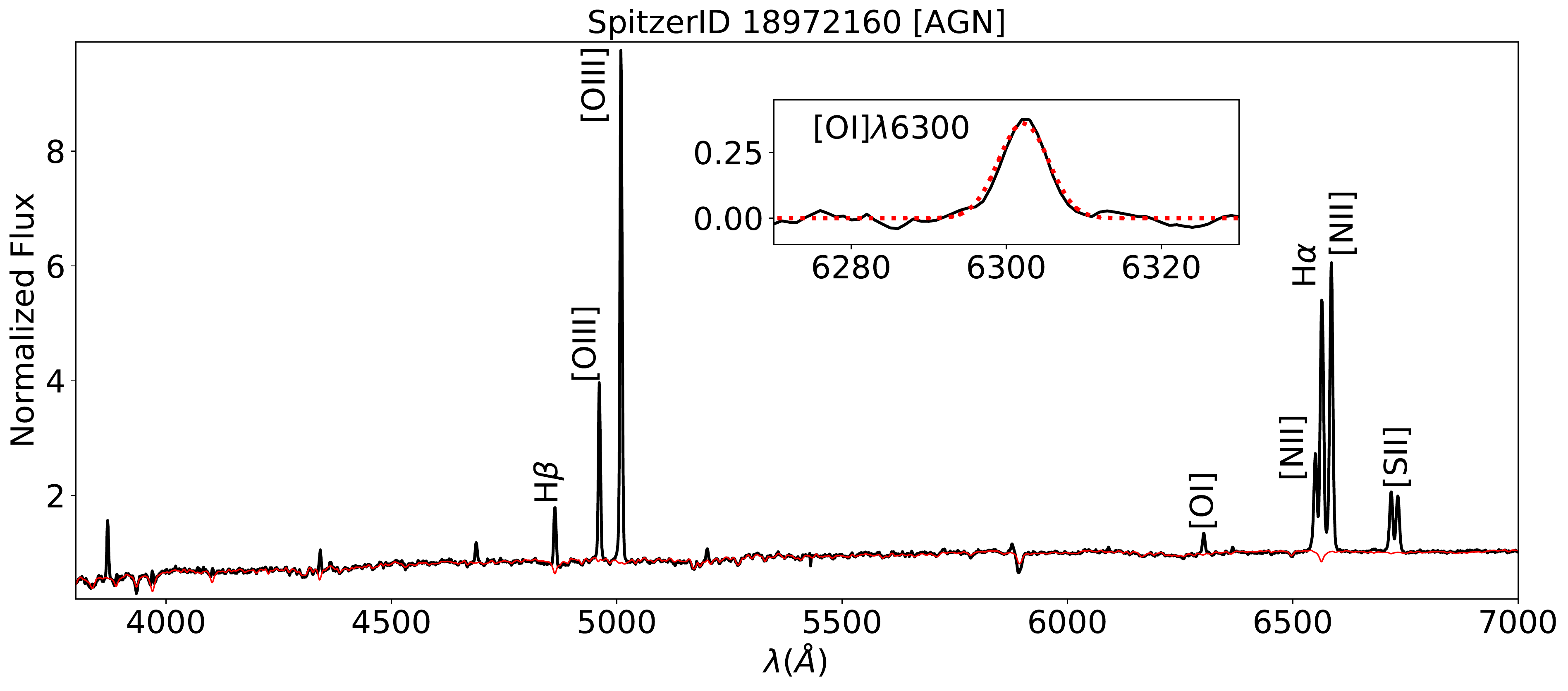}
\includegraphics[width=0.8\textwidth]{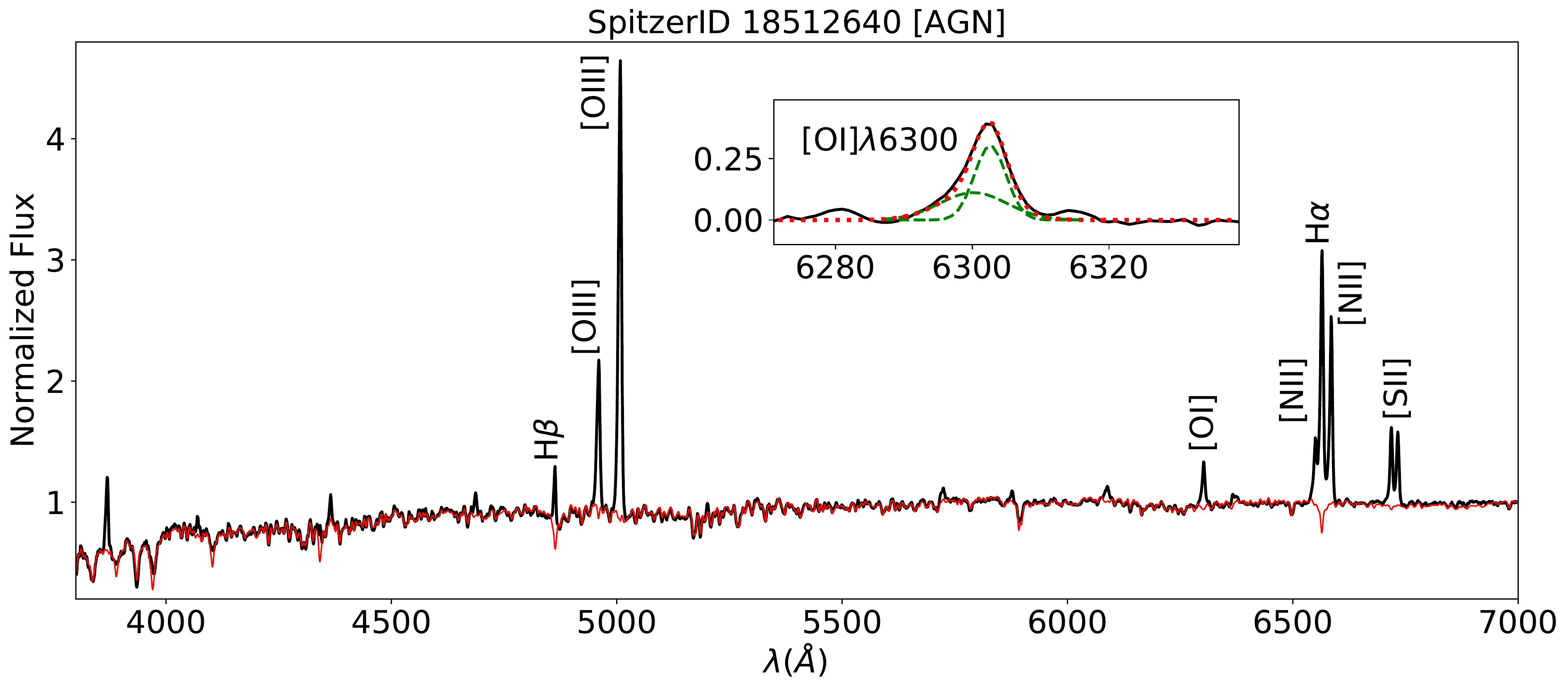}
\includegraphics[width=0.8\textwidth]{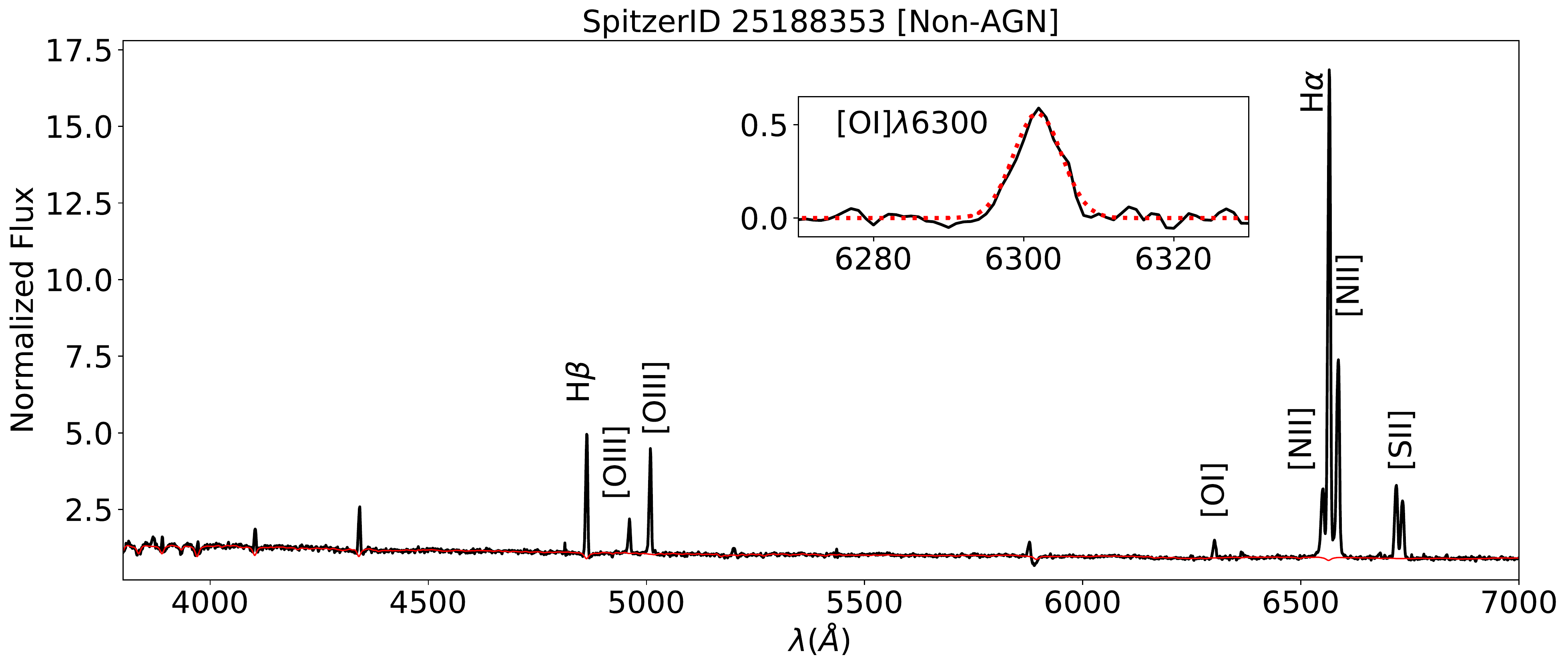}
  \caption{Examples of SDSS spectra for three objects of sample Y. In red we show the synthesized spectra from {\sc starlight} code and the inserts show a zoom in the  [O\,{\sc i}]$\lambda6300$ emission line region from the stellar component subtracted spectra. The resulting fits of the [O\,{\sc i}] profiles obtained with {\sc ifscube} code are shown as dotted red curves and in the middle panel, where the profile is better fitted by two Gaussians, the individual components are shown as dashed green lines. The fluxes are normalized by their values at 5700\,\AA\ and the spectra are corrected to the rest frame.
 }
  \label{fig:spectra}
\end{figure*}

  In addition, for 27 objects the H$_2$S(3)\,9.665\,$\mu$m line was detected, but with no detection of PAH\,11.3\,\,$\mu$m emission lines. Only one galaxy is in the SDSS database.  Table\,\ref{tab:sample} lists the number of galaxies of each sub-sample as well as the number of objects with SDSS data available.  In Figure~\ref{fig:spectra} we present examples of the typical SDSS spectra.  
  
  The flux limits of the Spitzer data may introduce biases in the H$_2$/PAH line ratios in each sample compared to a volume--limited sample of galaxies. However, the key scientific results or our paper focus on the objects in the sample Y with the strongest H$_2$ emission, which are the least affected by the biases, and therefore our main results are not affected.

%\subsection{Emission-Line Fluxes and Kinematics}

\subsection{Stellar population synthesis}

%\textcolo{red}{ In order to have a pure emission line spectrum, we need to remove the stellar underlying features from the gas emission. Therefore, one can fit the stellar continuum with combinations of simple stellar populations (SSPs).} 
The stellar population synthesis is performed using the {\sc starlight} code, which is described in \citet{cid04,cid05}. {\sc starlight}  fits the observed spectrum $O_{\lambda}$ with a model spectrum $M_{\lambda}$ obtained as a linear combination of $N_{\star}$ SSPs. The fitting result is a final population vector {\bf x}, whose components represent the fractional contribution of each SSP to the total synthetic underlying flux at wavelength  $\lambda_0$ \citep{cid04,cid05}. Extinction $A_V$ is modelled as due to a foreground dust layer with the wavelength dependence from \citet{cardelli89}. The full model is: \\

\begin{equation}
{\it M_{\lambda}= M_{\lambda 0} \left[ \sum_{j=1}^{N_{\star}} x_j b_{j, \lambda} r_{\lambda} \right] \otimes G(v_{\star}, {\sigma}_{\star})},
\end{equation}
where $M_{\lambda 0}$ is the synthetic flux at the normalisation wavelength and $G(v_{\star}, {\sigma}_{\star})$ is a Gaussian function used to model the line-of-sight stellar velocity distribution centered at velocity $v_{\star}$ with dispersion $\sigma_{\star}$. The term $x_j$ is the {\it j}th population vector component of the base of elements, defined as $b_{j, \lambda}$. All spectra of the SSP as well as the observed data are normalized to  unity at $\lambda_0$, so that the reddening term is $r_{\lambda}=10^{-0.4(A_{\lambda}-A_{\lambda 0})}$. The final fit is carried out through a $\chi^2$ minimization procedure.

The base set, e.g. the SSPs used in the fits, are those in the standard {\sc starlight} distribution and were taken from \citet{bc03} models.  These models provide an adequate spectral and age resolution to fit our data and are widely used in the study of stellar populations in nearby galaxies, which makes the comparison of our results with those of the literature straightforward. In addition, the {\sc starlight} code is optimized to run with \citet{bc03} models. The base set and fitting range used are described in \citet{mallmann18}. It comprises 45 SSPs with 15 ages (0.001, 0.003, 0.005, 0.010, 0.025, 0.040, 0.101, 0.286, 0.640, 0.905, 1.43, 2.50, 5.00, 11.00 and 13.00 Gyr) and three metallicities (0.1, 1 and 2.5 Z$\odot$). To allow for an AGN component, we also add a featureless component to the base set, represented as a power law function of the form $F_\lambda \propto \lambda^{-0.5}$ \citep[e.g.][]{cid05}. The fitting range is between 3800\,\AA\ to 7000\,\AA\ with normalization point $\lambda_0  = 5700$\,\AA. We present examples of the fitting in Fig.~\ref{fig:spectra}.
%As extinction law we used the one presented by \citet{cardelli89}. 

%The robustness of the fits can be measured by the \str\ output parameter ADEV, which is the average deviation over the fit of pixels $|O_{\lambda}-M_{\lambda}|/O_{\lambda}$.

\subsection{Emission line fluxes and kinematics}

\begin{figure}
  \centering
  \includegraphics[width=0.49\textwidth]{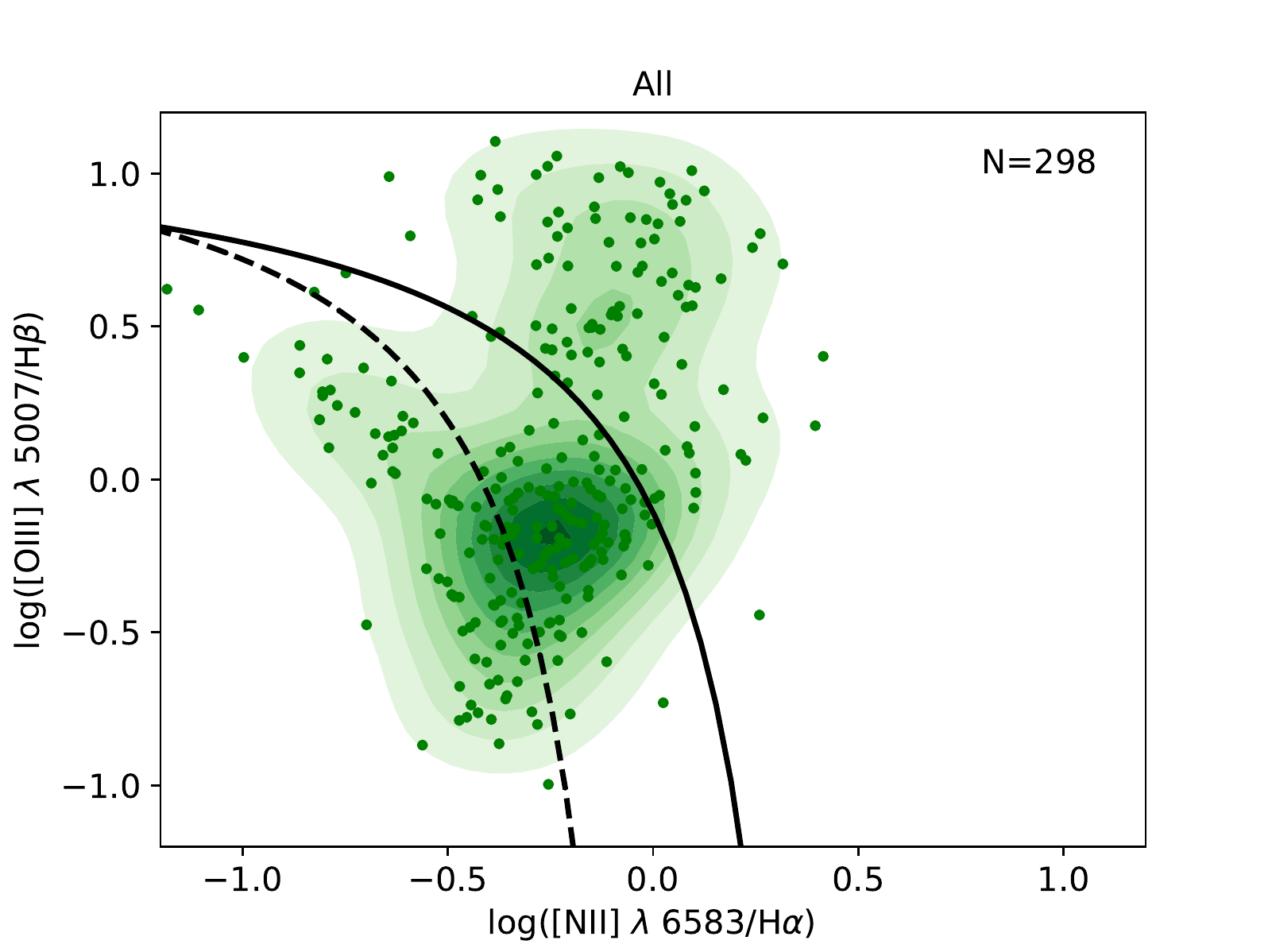} 
  \includegraphics[width=0.49\textwidth]{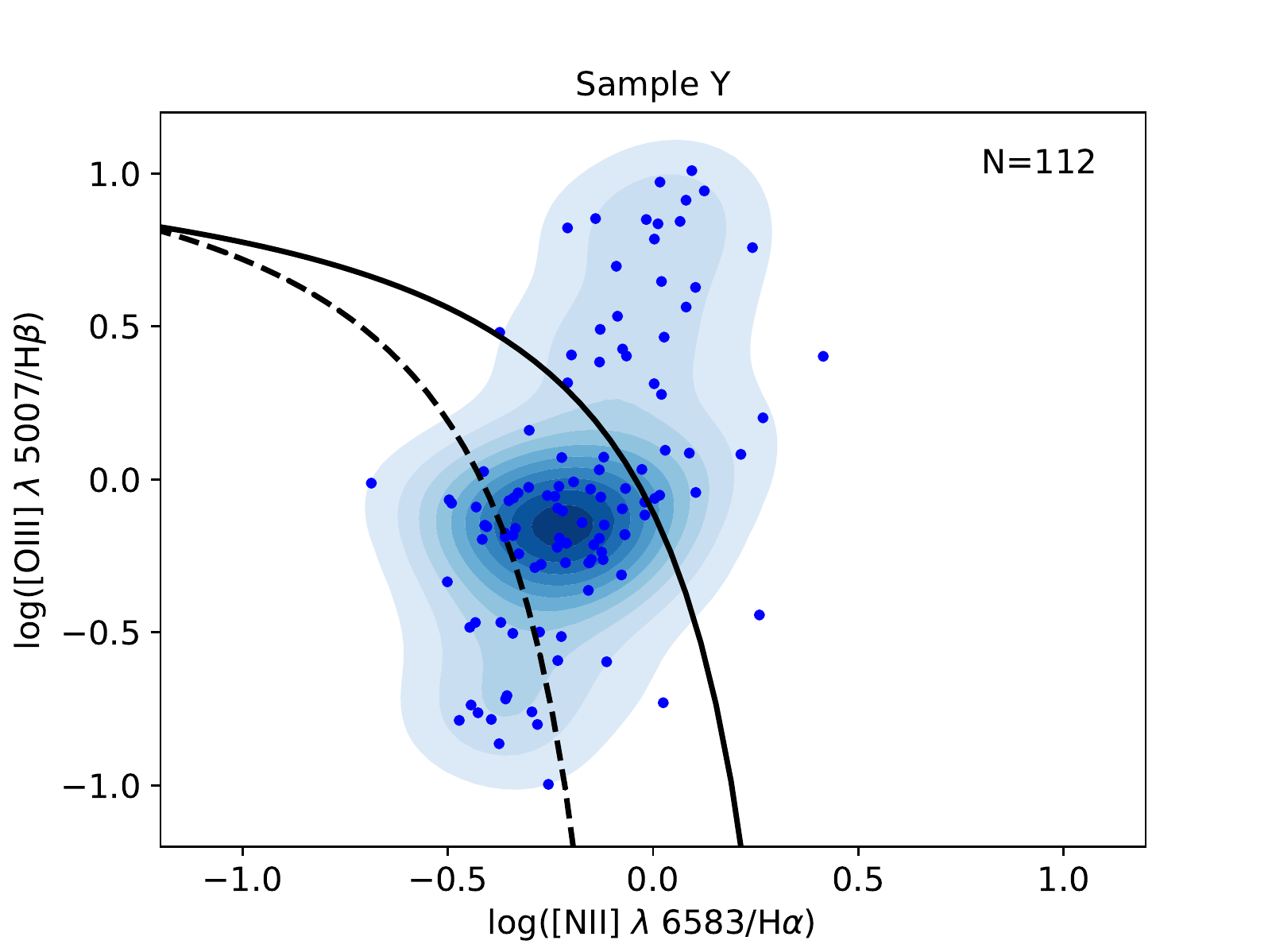} 
\includegraphics[width=0.49\textwidth]{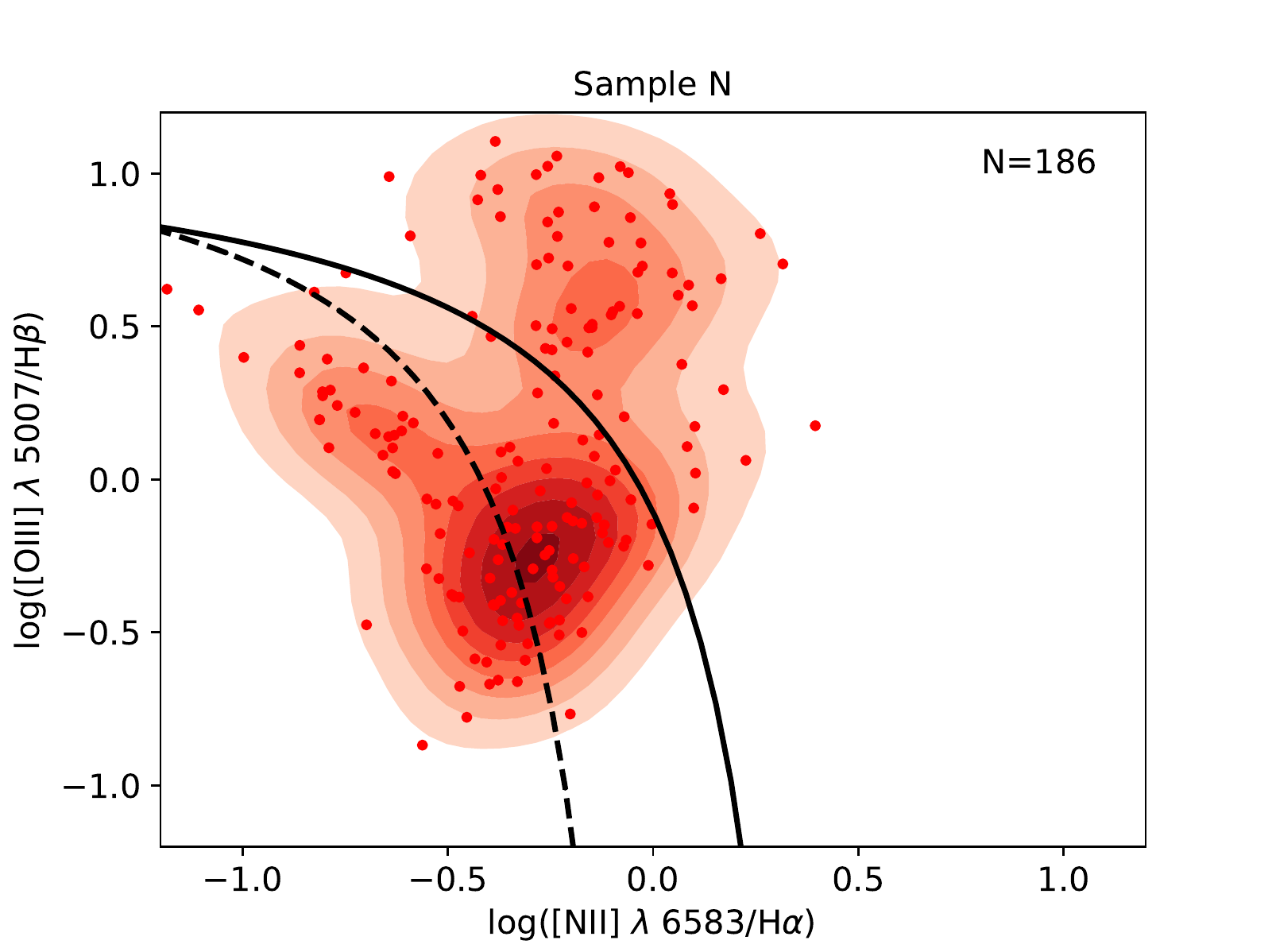} 

  \caption{BPT diagrams for the galaxies of the Spitzer sample with data available in the SDSS. The top panel shows the whole sample, the middle panel is for the Sample Y and the bottom panel for the Sample N.  The continuous line is from \citet{kewley01} and the dashed line from \citet{kauffmann03}. The  number of points ($N$) used in each plot is shown in the top left corner of the corresponding panel.  It is smaller smaller than the size of each sample, as for some objects at least one emission-line flux was not available.}
  \label{fig:bpts}
\end{figure}

 Fluxes of  H$_2$S(3)\,9.665\,$\mu$m and PAH\,11.3\,\,$\mu$m  features measured from Spitzer spectra are available from \citet{lambrides}.  These fluxes are aperture-corrected by flux-calibrating Spitzer spectra against WISE \citep{wright10} fluxes. Due to the low spectral resolution, no kinematic information of these features is available in Spitzer data.

The best-fit parameters of optical emission lines are from \citet{thomas13}, who fit the galaxy spectra using the Penalized Pixel-Fitting \citep[pPXF; ][]{cappellari04,cappellari17} and Gas AND Absorption Line Fitting \citep[GANDALF; ][]{sarzi06,oh11} codes to derive the stellar kinematics and emission line properties. During the fitting, the authors adopt the stellar population models from \citet{maraston11} to represent the continuum/absorption spectra and each emission-line profile is fitted by a single-Gaussian component. 

In order to verify if the distinct fitting and SSP models used by \citet{thomas13} and our methods result in similar measurements for the emission lines,  we use the {\it IFSCube} python package\footnote{https://ifscube.readthedocs.io} to fit the emission-line profiles seen in the residual spectra. The residual spectra are obtained by the subtraction of the continuum/absorption spectra modeled by the {\sc starlight} code from the observed spectra. We fit all spectra using a single-Gaussian and double-Gaussian functions per line.  However, only for $\sim$5 per cent (14 objects) of our sample, we find that the emission-line fluxes and the velocity dispersion of the broad component exceed 20 per cent of the fluxes and velocity dispersion of the narrow component. The middle panel of Fig.~\ref{fig:spectra} shows an example spectrum, which fulfills these criteria. Since the difference between \citet{thomas13} and our measurements is small, we use the emission-line parameters derived from the single-Gaussian fit throughout this paper.

We compare our measurements for [O\,{\sc iii}]$\lambda5007$/H$\beta$ and [N\,{\sc ii}]$\lambda6583$/H$\alpha$ from the single-Gaussian fit with those of \citet{thomas13} and find that they are very similar, with a mean difference of 0.07 dex for the first ratio and 0.008 dex for the latter. %Thus, we conclude that our measurements are consistent with those of \citet{thomas13}.
In addition, the {\it IFSCube} does not provide reliable estimates for the uncertainties. Thus, we  use the measurements from \citet{thomas13}, as they are easily available through the CasJobs server\footnote{http://skyserver.sdss.org/casjobs} (making our work easy to reproduce) and have been extensively used \citep[e.g.][]{rembold17}.

\section{Results}

We use the physical parameters of the optical and mid-infrared emission lines to characterize our sample and to investigate the relation between the molecular gas emission and the ionized gas excitation and kinematics. In Sec.~\ref{sec:sub-samples} we compare  samples Y and N in terms of the optical properties. Sec.~\ref{sec:relH2Opt} presents the relation between the molecular and ionized gas emission, while in Sec.~\ref{sec:relH2SP} we compare the stellar population and molecular gas properties and Sec.~\ref{sec:relH2shocks} investigates the origin of the molecular gas emission. Unless specified, we use all points of each plot to compute the $P_{\rm rank}$ values throughout this section.

\subsection{Comparing the sub-samples}\label{sec:sub-samples}

\begin{figure*}
  \centering
\includegraphics[width=0.49\textwidth]{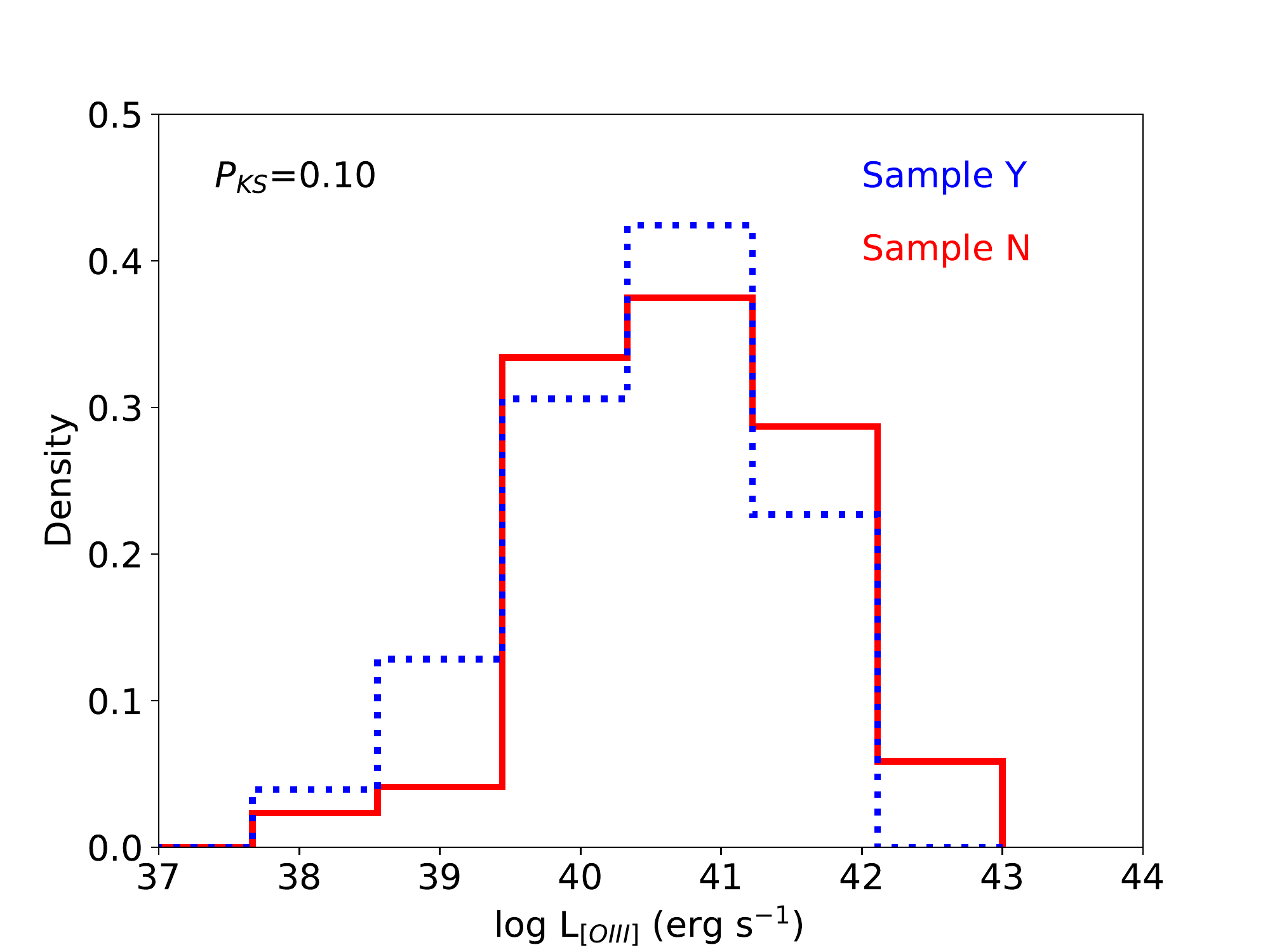}
\includegraphics[width=0.49\textwidth]{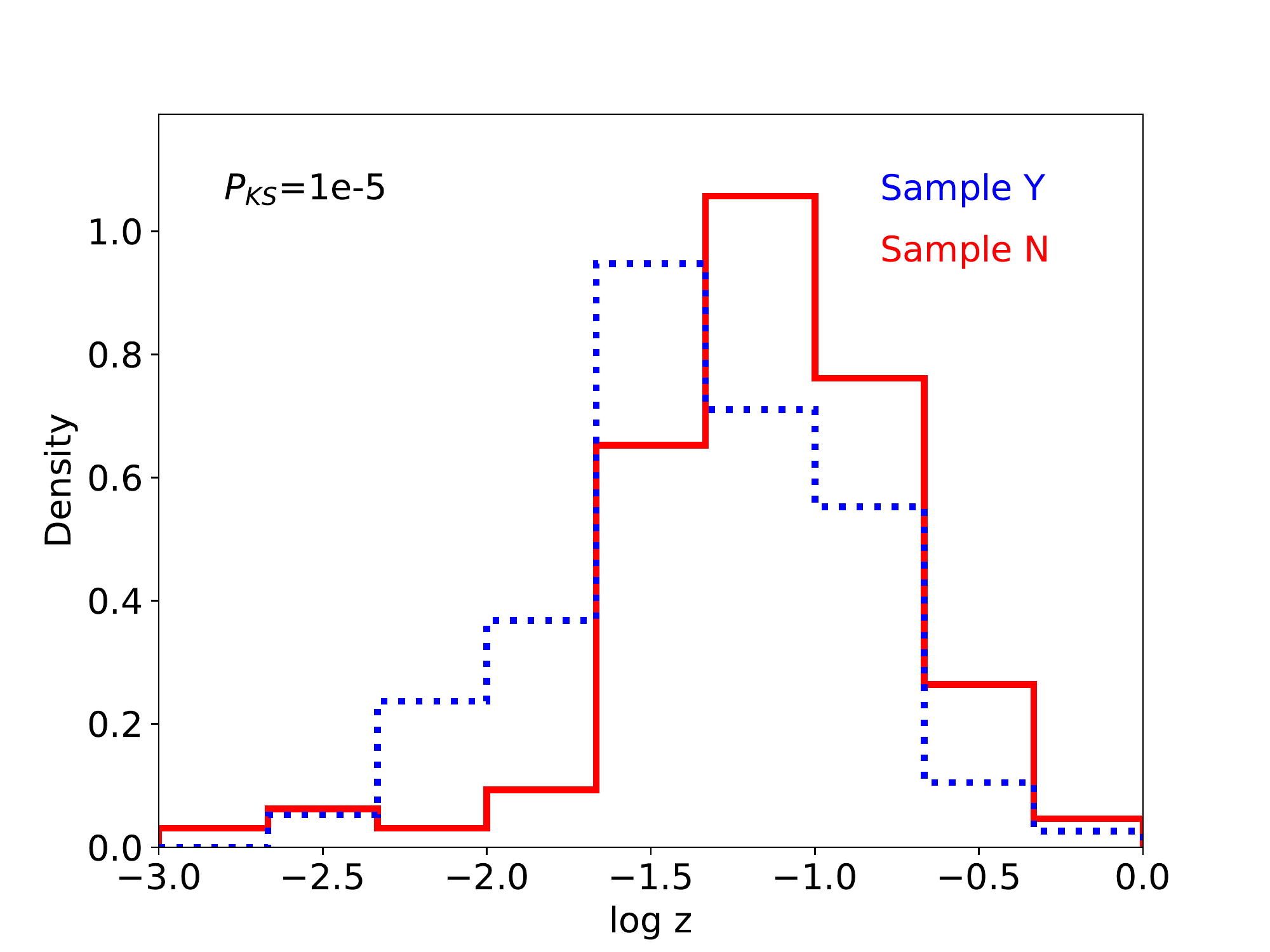}
\includegraphics[width=0.49\textwidth]{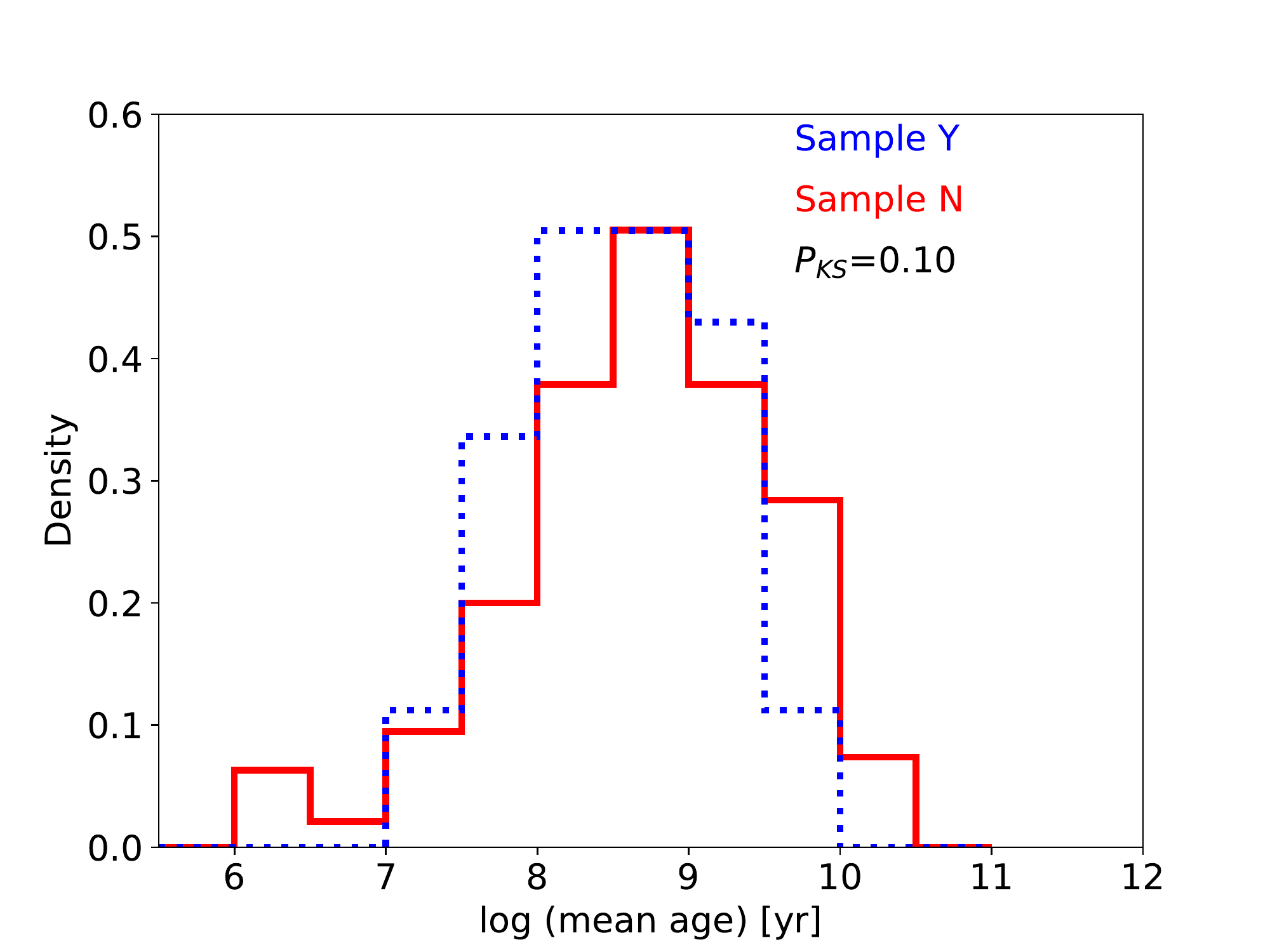}
\includegraphics[width=0.49\textwidth]{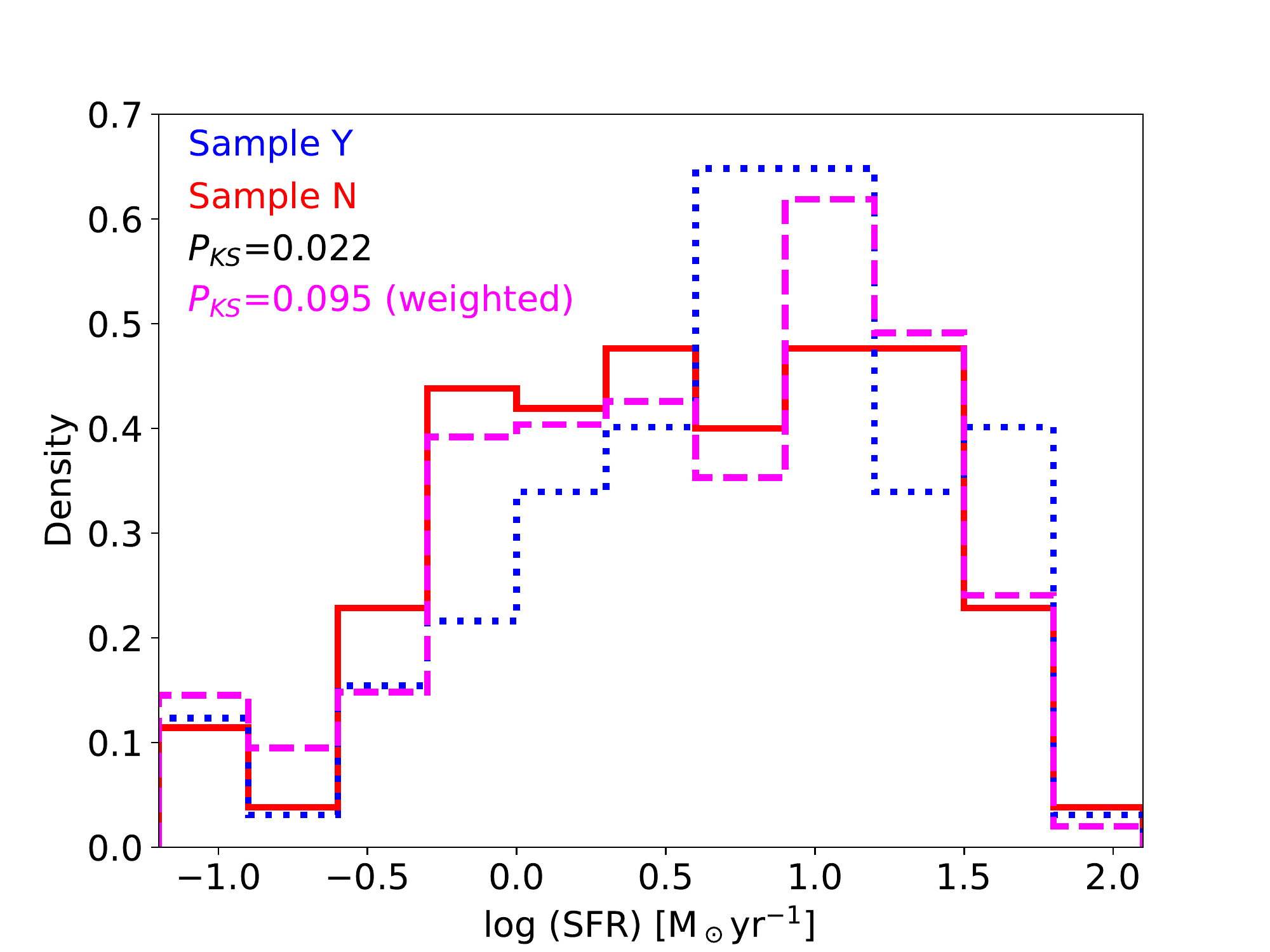}
  \caption{Comparison of the samples Y and N in terms of the [O\,{\sc iii}]$\lambda5007$ luminosity (top left), redshift (top right), mean age of the stellar populations (bottom left) and $SFR$ during the last 10 Myr (bottom right). The $P_{KS}$ values are shown in each panel. The green dashed line shows the distribution of SFR of the sample N, weighted by the redshift distribution (see text).}
  \label{fig:hist}
\end{figure*}

In Figure~\ref{fig:bpts} we present the [N\,{\sc ii}]$\lambda$6583/H$\alpha$ vs. [O\,{\sc iii}]$\lambda$5007/H$\beta$ emission-line ratio diagnostic \citep[BPT; ][]{bpt81} diagrams for all galaxies with data available in the SDSS archive (top panel), for the sample Y (middle panel) and for the sample N (bottom panel). Objects in the upper right part of the diagram are thought to be dominated by AGN photo-ionization, and objects in the bottom left dominated by ionization typical of star-forming  galaxies. 
%The continuous line is from \citet{kewley01} and delineate the region with ionization dominated by AGN (points that lie right and above the line) and the dashed line is from \citet{kauffmann03}. Points below and left to this line represent pure star-forming (SF) galaxies.   

As we see from the color density contours in the BPT diagrams, most objects are in the central region, indicating a combination of AGN-dominated and star-forming-dominated ionization. We use the BPT diagram to discriminate our sources as AGN and non-AGN throughout this paper. We consider all objects that lie above and right the line of \citet{kewley01} to be AGN-dominated.
%and have H$\alpha$ EqW greater than 3\,\AA. The latter criteria is useful to discriminate between ionization by AGN and post-AGB stars, which show H$\alpha$ EqW$<$3\,\AA~  \citep{cid10}.  
The sample Y (sample N) is composed by 31.3\,\% (32.2\,\%) of AGN,  22.3\,\% (38.2\,\%) of star-forming galaxies and 46.4\,\% (29.6\,\%) of transition objects. 

The equivalent width of the PAH\,6.2\,$\mu$m feature ($EW[PAH]$) can be used as an indicator of AGN \citep{laurent00,peeters04,brandl06,sales10,zakamska16a}. In galaxies where the AGN contribution to the mid-infrared emission is larger than 50 per cent, the $EW[PAH]$ is usually smaller than 0.27\,$\mu$m, while transition objects and star-forming galaxies show $EW[PAH]>0.27$\,$\mu$m  \citep[e.g.][]{lambrides}. By comparing the BPT and $EW[PAH]$ based AGN classification in our sample, we find that 40 per cent of the optically selected AGN show $EW[PAH]<0.27$\,$\mu$m and about 65 per cent of the objects classified as AGN using the $EW[PAH]$ are also classified as AGN in the optical. Such discrepancies among distinct AGN classification methods are well known \citep[e.g.][]{heckman14}. 
%In this paper, we use the BPT classification for distinguishing AGN from non-AGN sources. 

%All BPT diagrams show most points in the region between the Kauffman's and Kewley's lines (as seen from the colour density contours), indicating a composite ionization source, including AGN and stellar radiation. In addition, all samples present significant fractions of points in the AGN andstar-forming regions. No clear difference is seen in the BPT diagrams for the sample Y and sample N, suggesting that the detection or absence of the mid-infrared molecular lines is not only related to distinct power sources.

In Figure~\ref{fig:hist} we present the distributions of the [O\,{\sc iii}]$\lambda$5007 luminosity ($L_{\rm [OIII]}$), redshift ($z$), mean age of the stellar populations weighted by the light and star-formation rates ($SFR$) derived from the spectral synthesis using the {\sc starlight} code over the last 10 Myr.  The reported mean age is calculated following \citet{cid05}:
 \begin{equation}
\langle \log\,t_L\rangle \,= \frac{ \sum_{j=1}^{N_\star} x_j \, {\rm log}(t_j) }{ \sum_{j=1}^{N_\star} x_j },
\end{equation}
\noindent where t$_j$ is the age of the template $j$. 

Since our base spectra are in a proper unit of L$_{\odot}$ \AA$^{-1} M_{\odot}^{-1}$, and our observed spectra ($O_{\lambda}$) are in units of $\rm erg/s/cm^2/$\AA, the SFR over the last 10\,Myr can be computed assuming that the  mass of each base component (j) which has been processed into stars can be defined as:
\begin{equation}
    M_{\star,j}^{\rm ini} = M_{j}^{\rm ini} \times \frac{4\pi d^2}{L_\odot},
\end{equation}
where $M_{\star,j}^{\rm ini}$ is given in M$_\odot$, $M_{j}^{\rm ini}$ is a parameter computed by {\sc starlight} and related to the mass that has been converted into stars by j-th element and its flux. This parameter is given in ${ \rm M_\odot\,erg s^{-1} cm^{-2}}$ and $d$ is the distance to the galaxy in cm.
%and 3.826$\times10^{33}$ is the Sun's luminosity in erg\,s$^{-1}$. 
Thus, the SFR over the last years can be obtained by the equation:
\begin{equation}
    {\rm SFR} = \frac{\sum_{j_i}^{j_f} M_{\star,j}^{\rm ini}}{t_{j_f}-t_{j_i}}.
\end{equation}
To obtain SFR over the last 10\,Myr, we consider only the elements with ages $t\leq$10\,Myr (e.g. $j_f$=10\,Myr and $j_i$ = 0).

%}

We observe that the sample Y and sample N show similar distributions of $L_{[OIII]}$, suggesting that the presence or absence of rotational H$_2$ emission lines is not related to the power of the radiation field. The estimated $P_{\rm KS}$ confirms that the  $L_{[OIII]}$ distributions of both samples are statistically equivalent. The redshift distributions of the sub-samples are statistically distinct, as indicated by the small $P_{\rm KS}$. On average, galaxies from the sample N are located farther away than objects of the sample Y,  which could explain  the non-detection of weak molecular lines in the sample N in the farther objects.

The bottom panels of Fig.~\ref{fig:hist} show that the mean age of the stellar populations of the sample Y and sample N  are similar, whereas the distributions of SFRs are distinct  ($P_{\rm KS}=0.022$). 
%On the other hand, the distribution of recent $SFR$ are distinct with a confidence level larger than 95 per cent, as indicated by the K--S test ($p-value=0.05$ indicates 95\% of confidence level). 
The sample Y displays larger values of $SFR$. As the $SFR$ correlates with the amount of gas available to form stars, a possible interpretation of this result is that galaxies from the sample Y present a larger gas reservoir than objects of the sample N, suggesting that the detection of the molecular lines is closely related to the presence of molecular gas. 

However, the sample N is composed by objects, on average, farther away as compared to the sample Y. Therefore, the apparent difference in SFR could also be due to the fact that it is more difficult to properly measure the SFR for the  more distant sample N. In order to address this problem, we follow \citet{zakamska04}  and redshift-weight SFR distributions of samples Y and N for a more direct comparison. To this end, we divide both samples in 11 bins of $z$ (the same number of bins used to construct the SFR histograms)  and assign a weight $w$ to each object of the  sample N
\begin{equation}
    w=\frac{n_i N_Y}{m_i N_N},
\end{equation}
where $n_i$ and $m_i$ are the number of objects from samples Y and N in each redshift bin, respectively. $N_Y$ and $N_N$ are the total number of objects from each bin. The resulting weighted SFR distribution of the sample N is shown as a dashed green line in bottom right panel of Fig.~\ref{fig:hist} and indeed it is more similar to that of the sample Y. We compute $P_{\rm KS}=0.095$, indicating that there is no statistically significant difference between the SFR distributions in the sample Y and weighted sample N. 

In summary, the main differences between the samples Y and N are:  the sample Y composed by objects at smaller $z$,  shows a higher fraction of transition objects and a smaller fraction of star-forming galaxies in comparison with the sample N. Thus possible explanations for the non-detection of molecular lines in the sample N are that these lines may be too weak to be detected in the farther objects or they are related to the AGN physics, rather than to star formation. 

\subsection{The relation between molecular and ionized gas emission}\label{sec:relH2Opt}
%\textcolor{cyan}{I suggest try EW instead of line fluxes, the EW is telling something on the continuum emission too. And since EW are not sensitive to the flux calibration (at least less sensitive) you can try to make ratios from the infrared lines with optical ones.}
%\begin{figure}
%  \centering
%\includegraphics[width=0.49\textwidth]{Match_H2PAH_lum-eps-converted-to.pdf}
%  \caption{Plots of H$_2$S(3)\,9.665\,$\mu$m luminosity against H$_2$S(3)\,9.665\,$\mu$m PAH\,11.3\,\,$\mu$m luminosity. The purple star and orange triangle show the mean values for AGN  and non AGN, respectively.}
%  \label{fig:h2pahlum}
%\end{figure}

\begin{figure*}
  \centering
\includegraphics[width=0.48\textwidth]{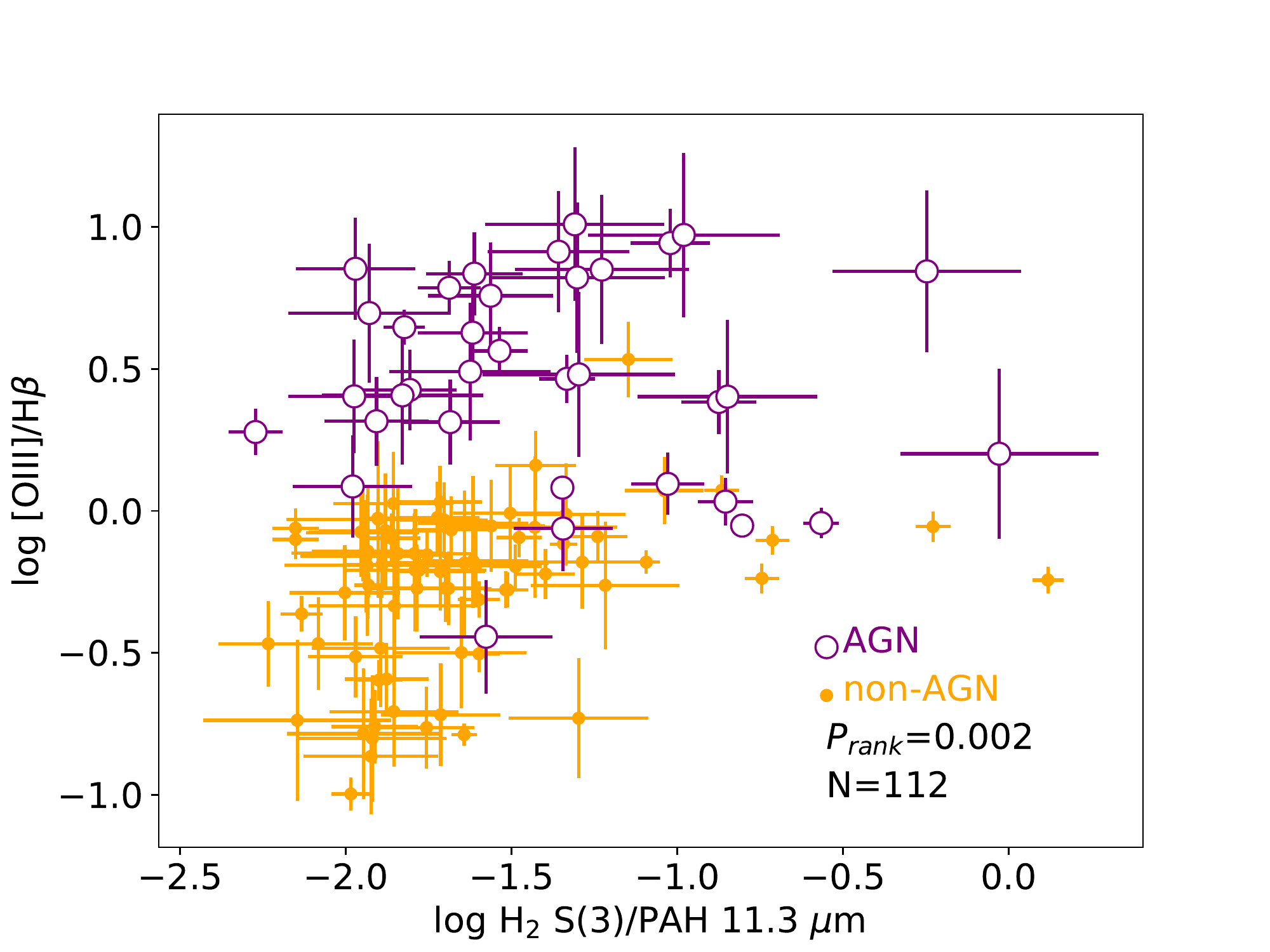}
\includegraphics[width=0.48\textwidth]{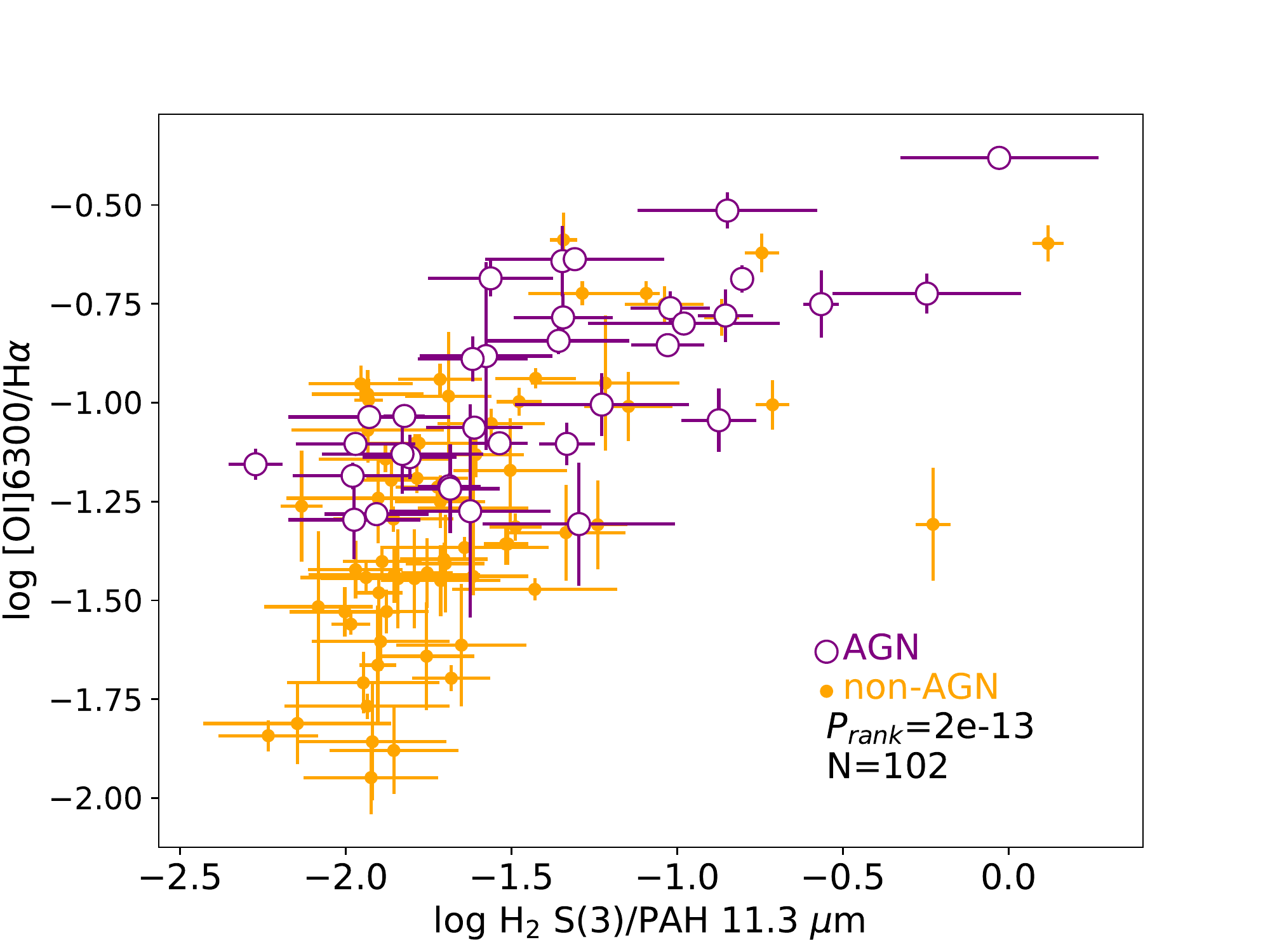}
  \caption{[O\,{\sc iii}]$\lambda5007$/H$\beta$ (left panel) and  [O\,{\sc i}]$\lambda6300$/H$\alpha$ (right panel) vs. H$_2$  S(3)\,9.665\,$\mu$m/PAH\,11.3\,\,$\mu$m emission-line ratios. AGN are shown as open circles and non-AGN as filled circles. The $P_{\rm rank}$ values are indicated in each plot. }
  \label{fig:ratio}
\end{figure*}

\begin{figure*}
  \centering
\includegraphics[width=0.49\textwidth]{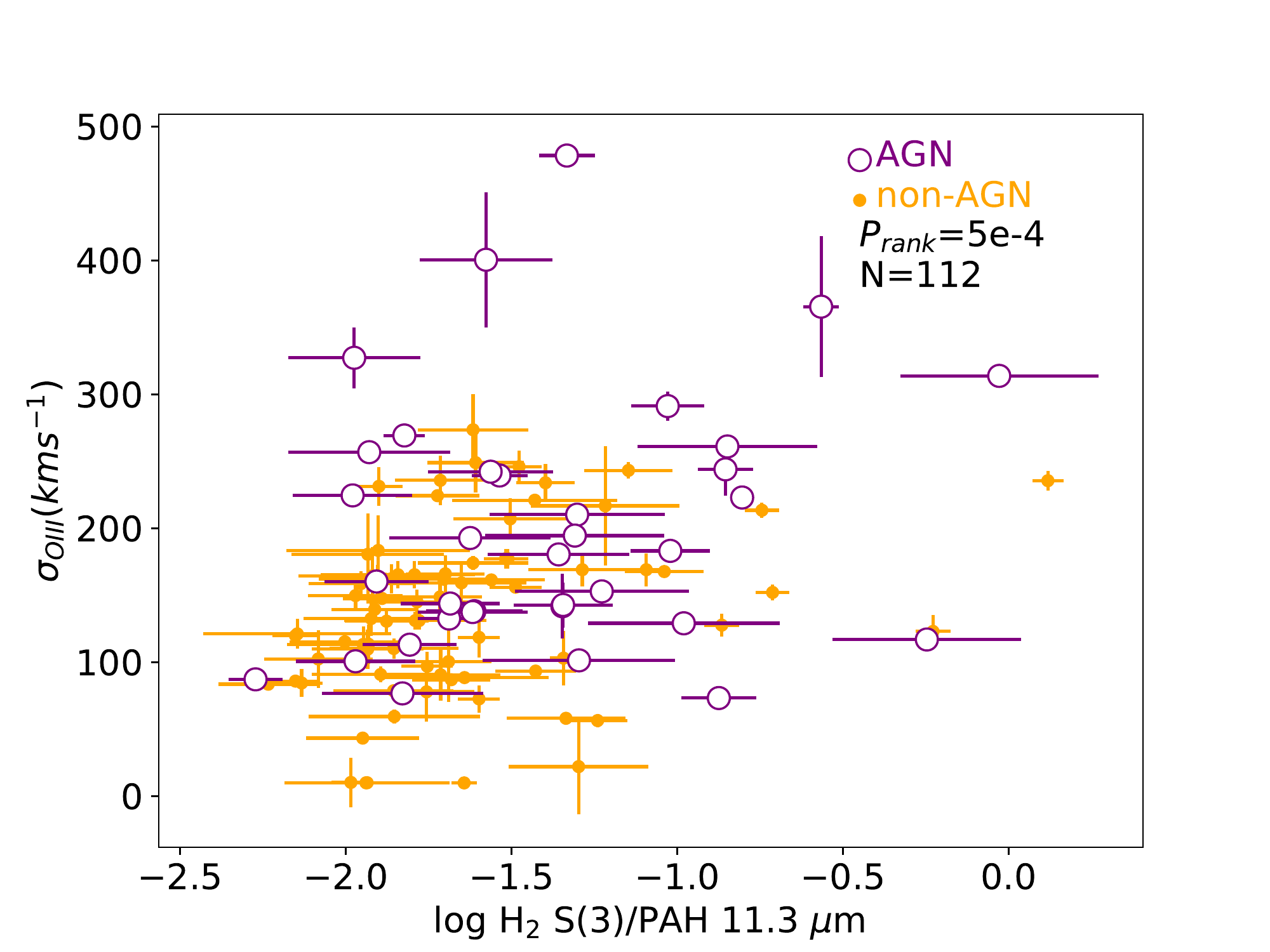}
\includegraphics[width=0.49\textwidth]{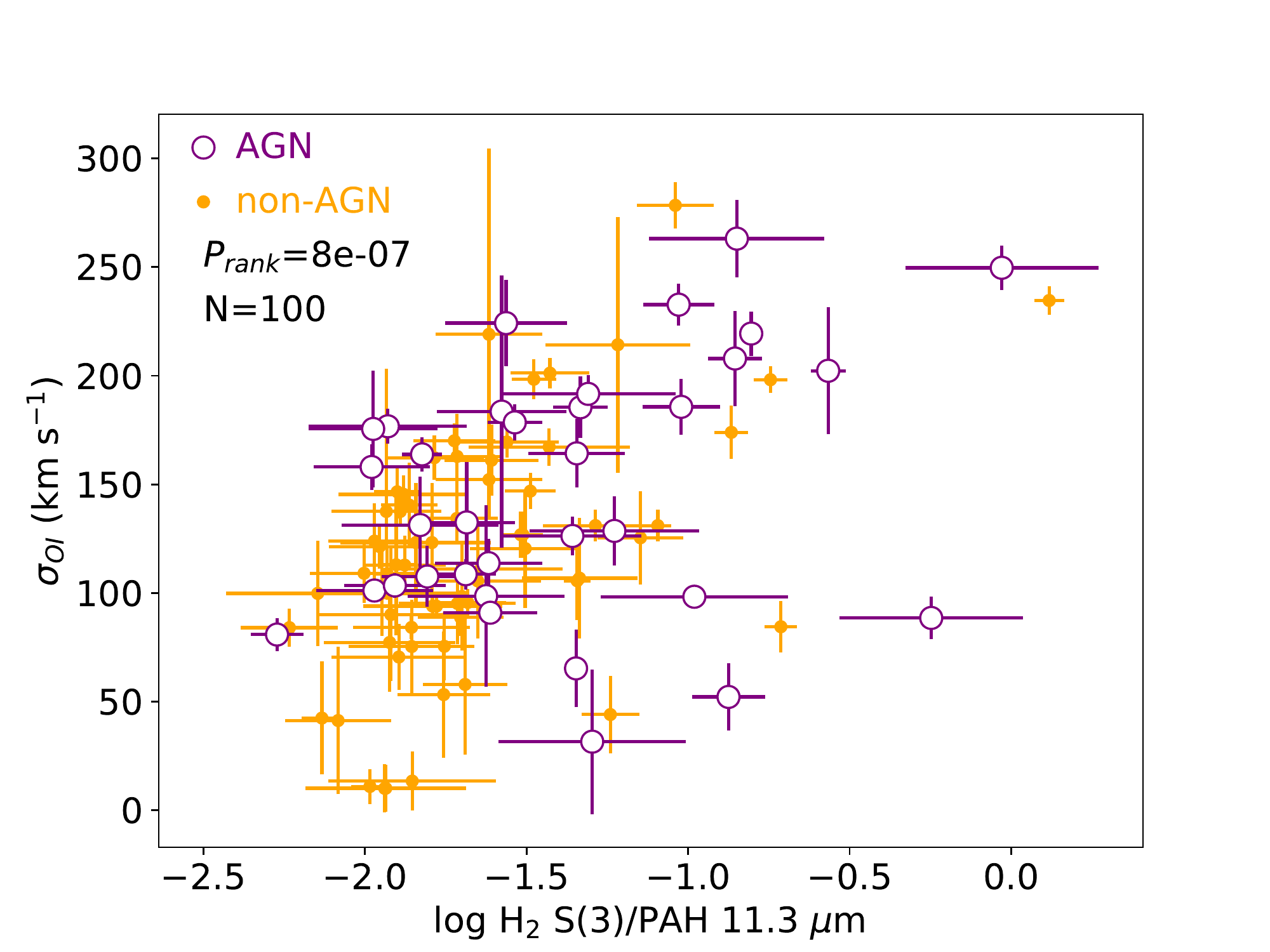}
  \caption{Plots of [O\,{\sc iii}]$\lambda5007$ (left panel) and  [O\,{\sc i}]$\lambda6300$ velocity dispersion vs. H$_2$  S(3)\,9.665\,$\mu$m/PAH\,11.3\,\,$\mu$m emission-line ratios. AGN are shown as open circles and non-AGN as filled circles. The $P_{rank}$ values are indicated in each plot.}
  \label{fig:sig}
\end{figure*}

In this section we explore the properties of the ionized gas in relation to the H$_2$S(3)\,9.665\,$\mu$m/PAH\,11.3\,\,$\mu$m emission-line ratio. For star-forming galaxies, the H$_2$/PAH ratio is approximately constant as both lines are produced in photodissociation regions \citep{roussel07}. 
%rom, figure\,\ref{fig:h2pahlum} we observe
We indeed find H$_2$S(3) and PAH luminosities correlate for the sample Y, and AGN show smaller values in both parameters. The larger typical PAH values in star-forming galaxies are likely due to a selection effect that starburst galaxies (with high SFRs) were more likely to be selected for follow-up Spitzer spectroscopy. Indeed, the median PAH$\lambda11.3\,\mu$m luminosity in the sample N (log\,(PAH) = 42.5) is slightly higher than in the sample Y (log\,(PAH) = 42.2).

We also find that the H$_2$/PAH ratio is larger in AGN and we refer to these higher values as ``H$_2$ excess''. In AGN hosts, an excess in the H$_2$ emission is observed relative to the PAH emission \citep{rigopoulou02,zakamska10,ogle12,stierwalt14,hill14,petric18,lambrides}.  We find average values for the H$_2$S(3)\,9.665\,$\mu$m/PAH\,11.3\,\,$\mu$m ratio of $(1.76\pm0.23)\times10^{-2}$ for star-forming galaxies, $(9.51\pm2.97)\times10^{-2}$ for AGN hosts and $(6.99\pm2.67)\times10^{-2}$ for transition objects. Thus, in AGN hosts we find that the H$_2$/PAH is about 5 times larger than in star-forming galaxies.
% *** Nadia: Wow, that's actually pretty striking... But how do we deal with upper limits in Figure 4? If H2 is not detected, then we don't include the objects, so aren't we biasing the AGN sample (which has lower PAHs) toward the high H2/PAH ratios?.. ***

Figure~\ref{fig:ratio} shows the relationships between  [O\,{\sc iii}]$\lambda5007$/H$\beta$ (left panel),  [O\,{\sc i}]$\lambda$6300/H$\alpha$ (right panel) and H$_2$S(3)/PAH. The [O\,{\sc iii}]$\lambda$5007/H$\beta$ is a tracer of the radiation field, while the [O\,{\sc i}]$\lambda$6300/H$\alpha$ is a tracer of shocks in neutral gas \citep[e.g., ][]{allen08,ho14}.  As indicated by the $P_{\rm rank}$ values, we find that both optical line ratios correlate with the H$_2$S(3)/PAH ratio, but a better correlation is found for  [O\,{\sc i}]/H$\alpha$, which may indicate that neutral gas shocks  play an important role in the production of the observed H$_2$ excess.

In order to determine if there is a kinematic signature of the shocks that may yield the H$_2$ excess, we plot the velocity dispersion ($\sigma$) of [O\,{\sc iii}]$\lambda$5007 and [O\,{\sc i}]$\lambda$6300 against  H$_2$S(3)\,9.665\,$\mu$m/PAH\,11.3\,\,$\mu$m. The corresponding plots are shown in Figure~\ref{fig:sig}. Both [O\,{\sc iii}]$\lambda$5007 and [O\,{\sc i}]$\lambda$6300 velocity dispersions are correlated with the H$_2$S(3)/PAH ratio, but as for the emission-line ratios, a better correlation is found for the [O\,{\sc i}] emission line.

\begin{figure*}
  \centering
\includegraphics[width=0.49\textwidth]{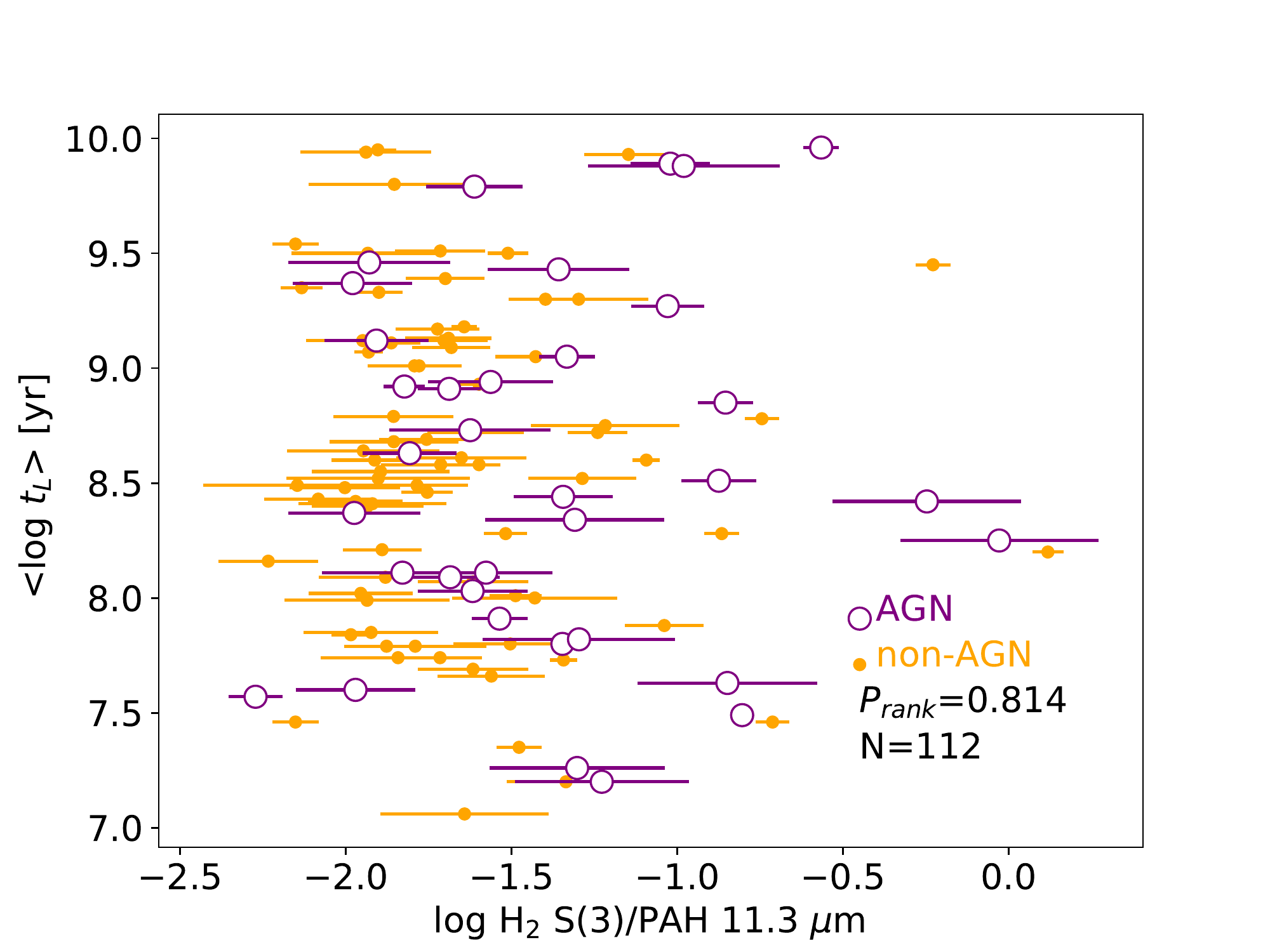}
\includegraphics[width=0.49\textwidth]{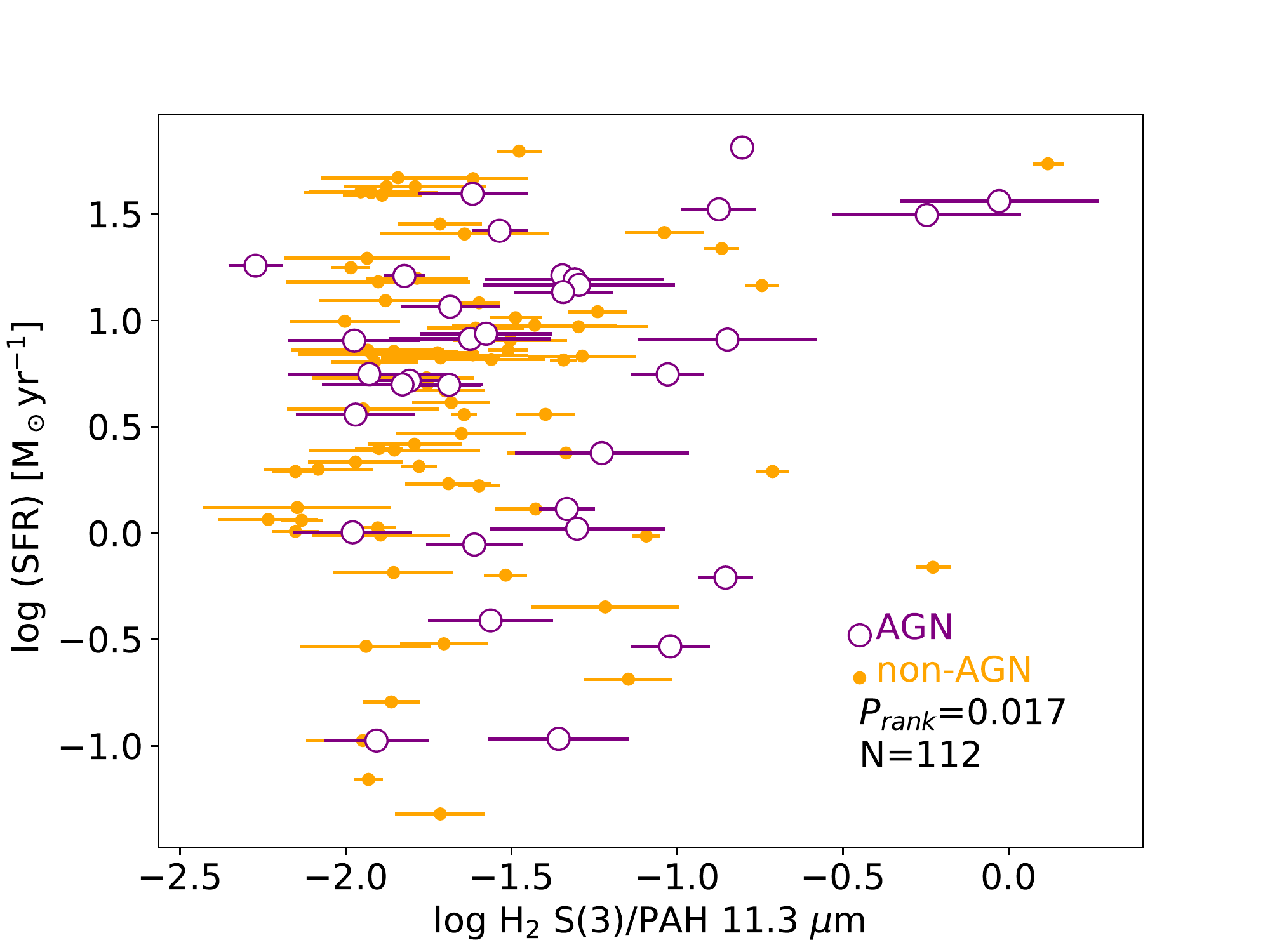}
  \caption{Mean age of the stellar populations (left) and SFR (right) vs. H$_2$  S(3)\,9.665\,$\mu$m/PAH\,11.3\,\,$\mu$m emission-line ratios. AGN are shown as open circles and non-AGN as filled circles. The $P_{\rm rank}$ value and the number of points are indicated in each plot.}
  \label{fig:sp}
\end{figure*}

\subsection{Molecular gas emission and stellar populations}\label{sec:relH2SP}

In the left panel of Figure~\ref{fig:sp} we show the mean age of the stellar populations, weighted by their contributions to the observed continuum emission, versus the  H$_2$S(3)/PAH emission-line ratio. As indicated by the derived $P_{\rm rank}$ values these parameters are not correlated. In the right panel, we use the SFR over the last 10 Myr instead of the mean age. The resulting $P_{\rm rank}$ value suggests that the parameters are correlated.% We find a correlation between SFR and H$_2$S(3)/PAH. 
%In agreement with the histograms presented in Fig.~\ref{fig:hist}, this correlation indicates that galaxies with a larger nuclear gas reservoir present larger SFR and H$_2$ excess.  
% *** Nadia: One sentence on what we are concluding from this observation? We are more likely to find excess molecular emission in objects that are currently actively star-forming?  ***

Previous studies found that the  H$_2$/PAH ratio is approximately constant for star-forming galaxies \citep{roussel07}, which is in apparent discrepancy with the correlation between SFR and H$_2$/PAH. %A detailed inspection of Fig.~\ref{fig:hist} show than the range of values H$_2$/PAH for non-AGN is much smaller than that for AGN. 
We find mean values of $\langle{\rm log\,H_2/PAH}\rangle=-1.37\pm0.08$ and $\langle{\rm log\,H_2/PAH}\rangle=-1.65\pm0.05$ for AGN and non-AGN, respectively. By computing the $P_{\rm rank}$ values between SFR and H$_2$/PAH, we do not find a statistically significant correlation for the non-AGN sample ($P_{\rm rank}=0.16$), while SFR and H$_2$/PAH correlates for the AGN sample ($P_{\rm rank}=0.04$). This indicates that the correlation seen for the whole sample is mainly due to the AGN, rather than the non-AGN sources. A possible interpretation for the correlation between SFR and H$_2$/PAH in the AGN sample is that the same gas that trigger the star formation is also triggers the AGN activity, connecting both processes \citep{perry85,terlevich85,norman88,cid01,n7582,mallmann18}. 

The absence of a correlation for non-AGN indicates that the origin of the $H_2$ emission excess is related to the AGN, rather than to star formation, in agreement with \citet{lambrides}. In addition, our results suggest that shocks due to AGN winds are present as indicated by the high [O\,{\sc i}] velocity dispersion in some of the objects in our sample. As expected, shocks contribute H$_2$ excitation over what is expected from star formation alone, which leads to the correlation between H$_2$/PAH and [O\,{\sc i}] velocity dispersion in Fig.~\ref{fig:sig}. Furthermore, the fact that there exists a correlation between S(3)/S(1) and the [O\,{\sc i}] velocity dispersion (Figure~\ref{fig:heat}) indicates that the H$_2$ excitation temperatures are coupled to the excitation mechanism and may be potentially used as shock diagnostics.
%In addition, our results suggest that shocks due to AGN winds are present as indicated by the correlation between H$_2$S(3)9.665\,$\mu$m/H$_2$S(1)17.03\,$\mu$m and [O\,{\sc i}]$\lambda$6300\AA\ velocity dispersion (Figure.~\ref{fig:heat}).
This leads to the conclusion that the way as AGN impact the interstellar medium is mainly due to mechanical feedback, instead of radiative feedback. % Although, most non-AGN present low  [O\,{\sc i}]  velocity dispersion, it should be noted some objects present high $\sigma$ values and for these objects, shocks due to stellar winds may also be important.

\subsection{Gas kinematics and molecular gas emission}\label{sec:relH2shocks}

%The origin of the mid-IR molecular hydrogen emission from  galaxies is not yet understood. The roles of AGN and star formation to the production of the observed emission are still not clear. 
%Recently,  \citet{lambrides} found that AGN-dominated sources present on average  200 K larger temperature than non-AGN.  This result suggests that the AGN is playing an important role in heating the interstellar medium of their hosts, but the exact mechanism of this heating is not understood.

In order to further investigate the impact of the AGN in the interstellar medium, in Fig.~\ref{fig:sig} we examine the H$_2$S(3)9.665\,$\mu$m/H$_2$S(1)17.03\,$\mu$m emission-line ratio against the [O\,{\sc i}]$\lambda$6300\AA\ velocity dispersion. The former is a tracer of the  H$_2$ excitation temperature. As this ratio increases, the temperature also increases. The latter can be tracing the gravitational potential of the galaxy but also can be an indicator of shocks. A way to determine whether the [O\,{\sc i}]$\lambda$6300\AA\ velocity dispersion is tracing the gravitational potential is by comparing it with the stellar velocity dispersion ($\sigma_\star$). The stellar velocity dispersions of the galaxies in our sample are available from  \citet{thomas13}. Following \citet{ilha19}, we quantify the differences between the stellar and [O\,{\sc i}] velocity dispersions using the parameter $f_\sigma$:
\begin{equation}
f_\sigma = \frac{\sigma_{\text{[OI]}} - \sigma_{\star}} {\sigma_{\star}}.
\end{equation}
Higher values of $f_\sigma$ are indicative of a disturbed kinematics (e.g., the gas motions are inconsistent with the gravitational potential of the galaxy) and most probably due to outflows. This is further supported by the fact that for 70 per cent (10/14) of the objects that show emission lines with more than one kinematic component, the broad component is blueshifted by a few tens of km\,s$^{-1}$. Excess blueshift is a classical signature of outflows \citep{whittle85} since the receding redshifted part of the outflow tends to have a greater extinction than the blueshifted part. Similar disturbed gas kinematics can also be produced by gas inflows towards the center of the galaxies, but this scenario is unlikely in our sample, as inflows are usually associated to low velocity dispersion gas \citep{thaisa19}.

% *** Nadia: So, how do we know this? You previously said that H_2 inflows are common. In the handful of objects that have two-Gaussian components in [OI], do we see blueshifted components, redshift components, or equal likelihood? What about [OIII]? I know that [OIII] doesn't correlate as well with H2, but there are better data there. Can we tell if [OIII] is blue-shifted relative to the host galaxy frame? I think we should _prove_ that these are outflows, or at least use the limited available data 

\citet{ilha19} report median values of $f_\sigma$ for AGN and inactive galaxies of  $0.04$ and $-0.23$, respectively, based on measurements for the [O\,{\sc iii}]$\lambda$5007\AA\ instead of [O\,{\sc i}]$\lambda$6300\AA. They conclude that the higher values seen for AGN are due to gas outflows. 
For our sample, we find that AGN and non-AGN have similar $\sigma_\star$ distributions ($P_{\rm KS}=0.93$) and $\langle f_\sigma\rangle=0.24\pm0.07$ for AGN, $0.13\pm0.06$ for transition objects and $-0.24\pm0.07$ for star-forming galaxies. These values indicate a contribution of shocks to the [O\,{\sc i}]$\lambda$6300\AA\ emission from AGN and transition objects, possibly due to AGN driven winds. 

\begin{figure}
  \centering
\includegraphics[width=0.49\textwidth]{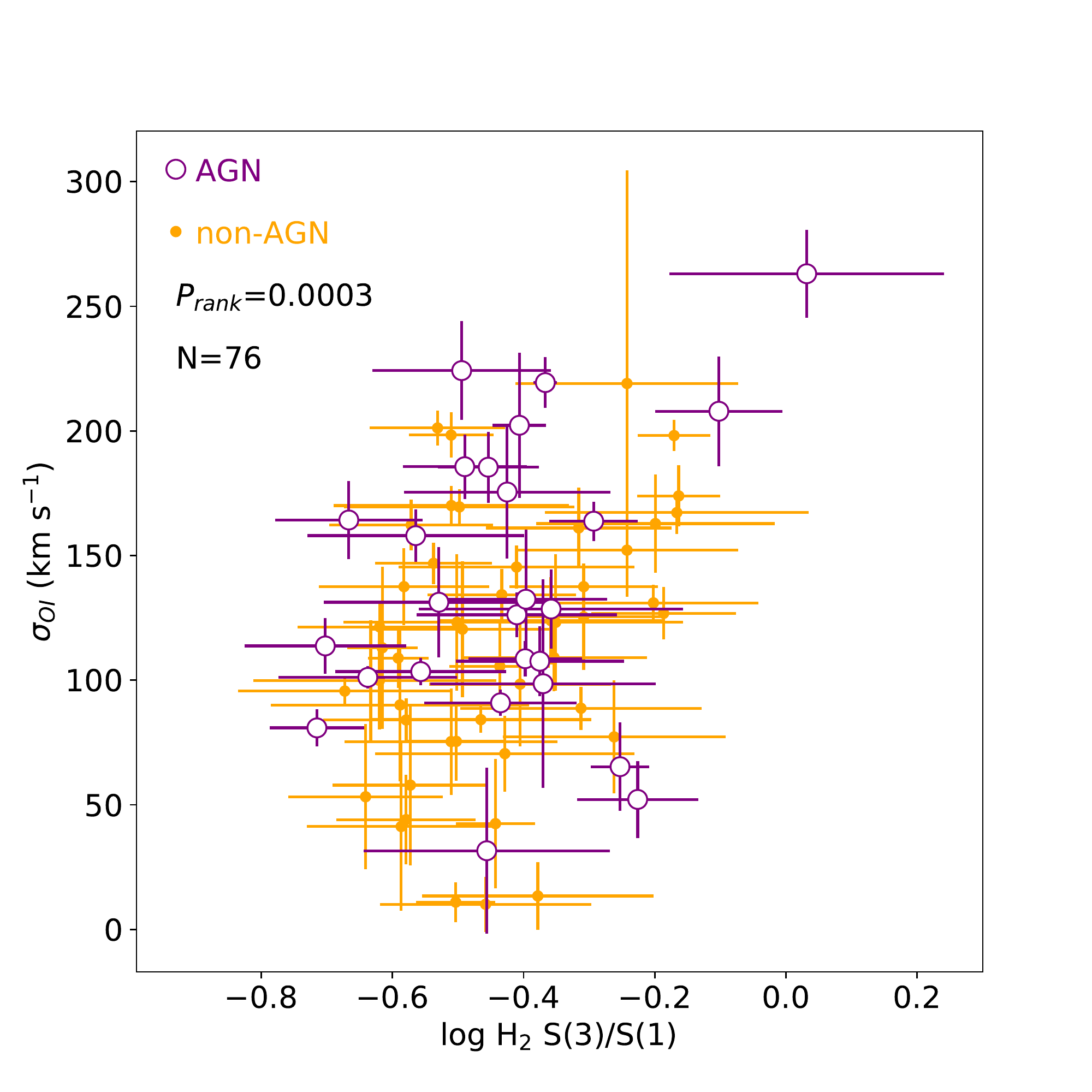}
  \caption{H$_2$S(3)9.665\,$\mu$m/H$_2$S(1)17.03\,$\mu$m emission-line ratio vs. the [O\,{\sc i}]$\lambda$6300\AA\ velocity dispersion. AGN (non-AGN) are shown as open (filled) circles, the number of objects and the $P_{\rm rank}$ value are indicated in the panel.}
  \label{fig:heat}
\end{figure}

Figure~\ref{fig:heat} shows the  H$_2$S(3)9.665\,$\mu$m/H$_2$S(1)17.03\,$\mu$m ratio against the [O\,{\sc i}]$\lambda$6300\AA\ velocity dispersion. Although the uncertainties in H$_2$S(3)9.665\,$\mu$m/H$_2$S(1)17.03\,$\mu$m ratio for individual sources are large, we find a correlation, with $P_{\rm rank}=3\times10^{-4}$. In addition, AGN present on average higher [O\,{\sc i}]  $\sigma$ values. This indicates that AGN play an important role in the production of the H$_2$ emission, supporting the results of \citet{lambrides}. A similar behaviour is observed if we plot $f_\sigma$ on the y-axis, but the uncertainties in $f_\sigma$ are high.

%In Figure~\ref{fig:sigo3o1}
We compare the [O\,{\sc iii}] and [O\,{\sc i}] velocity dispersions for the galaxies of our sample. As mentioned in Sec.~\ref{sec:sub-samples}, the detection of the molecular lines seems to be more related to the presence of a gas reservoir in the center of the galaxies, rather than to the radiation field. Thus, in the comparison of [O\,{\sc iii}] and [O\,{\sc i}] velocity dispersions, 
%in Fig.~\ref{fig:sigo3o1},
we include the whole sample, instead of only those galaxies with detected molecular lines.  A correlation is found between  [O\,{\sc iii}] and [O\,{\sc i}] $\sigma$ values, but the [O\,{\sc iii}] presents systematically higher velocity dispersion than [O\,{\sc i}]. The $f_\sigma$ values for the [O\,{\sc iii}] are also higher than those for [O\,{\sc i}]. A possible interpretation for this result is that the [O\,{\sc iii}] and [O\,{\sc i}] trace distinct phases of the outflow.

%In order to investigate the role of AGN and star formation in the production the observed H$_2$ emission excess seen in AGN hosts, we plot the [O\,{\sc iii}]$\lambda5007$ luminosity and SFR against the [O\,{\sc i}]$\lambda6300$ velocity dispersion, shown in Fig.~\ref{fig:winds}.

 We find that both $L_{[OIII]}$ and SFR  correlate with the [O\,{\sc i}]$\lambda6300$ velocity dispersion, in agreement with previous works \citep{yu19,woo17,ilha19}. In star-forming galaxies, gravitational instabilities alone cannot explain the observed gas velocity dispersion and stellar winds are required to produce the correlation between SFR and $\sigma$ \citep{yu19}.  Similar results are found for AGN, while non-AGN show lower values for both parameters \citep{woo17,ilha19}.% If we split the sample in AGN and non-AGN, we do not find correlations for AGN, while for the non-AGN sample, both SFR and [O\,{\sc iii}]$\lambda5007$ luminosity correlates with the [O\,{\sc i}] velocity dispersion.

%In order to investigate the role of AGN and star formation in the production the observed H$_2$ emission excess seen in AGN hosts, we plot the [O\,{\sc iii}]$\lambda5007$ luminosity and SFR against the [O\,{\sc i}]$\lambda6300$ velocity dispersion, shown in Fig.~\ref{fig:winds}. In star-forming galaxies, gravitational instabilities alone cannot explain the observed gas velocity dispersion and stellar winds are required to produce the correlation between SFR and $\sigma$ \citep{yu19}.  Similar results are found for AGN, while non-AGN show lower values for both parameters \citep{woo17,ilha19}. We find that both $L_{[OIII]}$ and SFR  correlate with the [O\,{\sc i}]$\lambda6300$ velocity dispersion. If we split the sample in AGN and non-AGN, we do not find correlations for AGN, while for the non-AGN sample, both SFR and [O\,{\sc iii}]$\lambda5007$ luminosity correlates with the [O\,{\sc i}] velocity dispersion.

%\begin{figure*}
%  \centering
%\includegraphics[width=0.47\textwidth]{sigma_OI_LOIII-eps-converted-to.pdf}
%\includegraphics[width=0.47\textwidth]{SFR_sigma_OI-eps-converted-to.pdf}
%  \caption{[O\,{\sc iii}]$\lambda5007$ luminosity (left) and SFR (right) vs. [O\,{\sc i}]$\lambda6300$ velocity dispersion. Open and closed symbols represent AGN and non-AGN, respectively, and the number of points and $P_{\rm rank}$ are indicated in each panel.}
%  \label{fig:winds}
%\end{figure*}

\section{Discussion}

%\citet{lambrides} find that AGN are responsible for an excess of H$_2$ emission in the mid-IR. We find that the AGN fraction in the sample of galaxies with detected H$_2$ emission (sample Y) is similar to that of those with no H$_2$ emission (sample N). However, the fraction of galaxies classified as transition objects (with contribution of both AGN and star-forming galaxies) in sample Y is larger that for the sample N. This result may indicate that indeed the AGN is playing an important role in the origin of the H$_2$ emission. On the other hand, we find that the SFRs seen for the sample Y are on average larger than those for sample N (Fig.~\ref{fig:hist}). In addition, we find a correlation between the SFR and   H$_2$S(3)\,9.665\,$\mu$m/PAH\,11.3\,\,$\mu$m  flux ratio (Fig.~\ref{fig:sp}). 

\subsection{Relationship between outflow phases}

Galactic outflows are a multi-phase phenomenon \citep{feruglio10,veilleux13,zakamska16b,eso,riffel19,gonzalez-alfonso17,shimizu19}, with different diagnostics suitable for the different phases. The relationship between the phases of the outflow is not well understood, and it is not yet known which phase carries most of the mass, momentum and energy of the outflow. Our paper addresses these important questions by examining the relationships between mid-IR diagnostics of star formation (PAHs) and warm molecular gas (rotational H$_2$ lines) on the one hand, and optical emission lines associated with neutral and ionized gas phases on the other hand. 

We find that both [O\,{\sc iii}]$\lambda5007$/H$\beta$ and  [O\,{\sc i}]$\lambda6300$/H$\alpha$ correlate with the H$_2$  S(3)\,9.665\,$\mu$m/PAH\,11.3\,\,$\mu$m line ratio (Fig.~\ref{fig:ratio}), but a much better correlation is found for the latter.  Similarly, we find a stronger correlation between H$_2$/PAH and the kinematics of [O\,{\sc i}] than we do with the kinematics of [O\,{\sc iii}]. Our findings suggest in galaxies with  H$_2$ excess, [O\,{\sc i}] and [O\,{\sc iii}] emission lines are emitted by gas which is not in dynamical equilibrium with the host galaxy. Additionally, because the correlations between H$_2$ and [O\,{\sc i}] are tighter than those between H$_2$ and [O\,{\sc iii}], we infer that the neutral and warm molecular gas phases are much more strongly coupled to each other than they are to the [O\,{\sc iii}]-emitting ionized gas. 

The same observations indicate that shocks are playing an important role in producing the H$_2$ emission. Indeed, H$_2$ is strongly correlated with [O\,{\sc i}], and [O\,{\sc i}]/H$\alpha$ is a known tracer of shocks \citep{monreal-ibero06,monreal-ibero10,rich11,rich14,rich15,ho14}. If the velocity dispersion of [O\,{\sc i}]  is larger than 150 km\,s$^{-1}$ and log\,[O\,{\sc i}]$\lambda6300$/H$\alpha \gtrsim -1.0$, shocks with velocities in the range of 160--300\, km\,s$^{-1}$ are the dominant excitation mechanism of the [O\,{\sc i}]. For smaller $\sigma$ and line ratio values, both shocks and photoionization contribute to the gas excitation \citep{ho14}.

%Considering that, in our sample, (i) the H$_2$  S(3)\,9.665\,$\mu$m/PAH\,11.3\,\,$\mu$m  correlates well with the [O\,{\sc i}] velocity dispersion and  [O\,{\sc i}]$\lambda6300$/H$\alpha$. (ii) The H$_2$ emission excess is associated to the presence of an AGN. (iii) The [O\,{\sc i}] velocity dispersion is not consistent with motions of gas subjected to the gravitational potential of the galaxies (as revealed by the comparison of the [O\,{\sc i}]  and stellar velocity dispersion in Sec.~\ref{sec:relH2shocks}). We conclude that shocks due to AGN driven outflows are responsible for the origin of the observed H$_2$ emission excess. 

Furthermore, not only are the shocks responsible for the  H$_2$ excess, but given the strength of the correlation between all measures of H$_2$ and [O\,{\sc i}] it suggests that the excess H$_2$ is produced in the same clouds as those that produce [O\,{\sc i}]. This is somewhat surprising because we normally think of neutral medium and dense molecular clouds as being two different components of the interstellar medium, and in particular star-forming molecular clouds in the Milky Way are much denser than the diffuse neutral component. Multiple numerical simulations demonstrated that dense molecular clouds would be very difficult to accelerate by an incoming wind \citep{klein94,scannapieco15,bruggen16,zhang17}. Instead such clouds would shred and become entrained in the wind.  Therefore, to explain the presence of recently discovered AGN-driven molecular outflows, theoretical models \citep{richings18a,richings18b} suggest that molecules can form within the already accelerated outflow. Possibly this is what we are seeing in both [O\,{\sc i}] and H$_2$.

%Considering that [O\,{\sc iii}] and [O\,{\sc i}] trace different gas phases, they may also be tracing distinct phases of the outflow. 

The better correlations found for [O\,{\sc i}] with the H$_2$/PAH, indicate that the H$_2$ and [O\,{\sc i}] emissions arise from similar outflow phases, while the [O\,{\sc iii}] originates from a higher velocity outflowing gas. This interpretation is consistent with results found for ULIRGs, that show a good correlation between  [O\,{\sc i}]$\lambda6300$/H$\alpha$ and H$_2$/PAH ratios, but the higher values of H$_2$/PAH seen in AGN cannot only be explained by the gas excitation due to the AGN radiation field, and shocks are necessary to explain the correlation \citep{roussel07,hill14}.

Our sample is composed mostly by low luminosity AGN ($L_{\rm bol}=10^{42-45}$ erg\,s$^{-1}$) and the kinematics suggest that most of the gas is in equilibrium with the galaxy and only a small fraction may be outflowing.  As the wind velocities from low-luminosity AGN are expected to be small, to disentangle the gravitational and wind components using single aperture spectra is not an easy task and still remains unresolved. We assume that there are no outflows in the non-AGN sample, and that the increased line widths in AGN  is due to the outflowing gas. Thus, we use the difference between the median line widths for AGN and non-AGN as a proxy of the velocity of the outflow ($v_{\rm out}$). We measure $v_{\rm out}^{\rm [OIII]}\sim100$\, km\,s$^{-1}$ for the [O\,{\sc iii}] and $v_{\rm out}^{\rm [OI]}\sim77$\, km\,s$^{-1}$ for the [O\,{\sc i}] outflow components, respectively.  
This is a very simplistic approach, which does not take into account possible differences in the mass distribution of AGN and non-AGN hosts in our sample. However, \citet{ilha19} find that AGN show higher values of gas velocity dispersion compared to inactive galaxies, matched by the AGN host properties, which include the morphological classification and stellar mass. They quantify this difference by the $f_\sigma$ parameter and interpret the higher values being due unresolved AGN ouflows. In addition, the derived mean outflow velocity in our sample is consistent with the values obtained from spatially resolved observations \citep{cresci15,kakkad16,slater19,diniz19}.  
%velocity dispersion for AGN ($\langle\sigma_{\rm AGN}\rangle$) and non-AGN ($<\langle\sigma_{\rm nAGN}>\rangle$) -- $v_{\rm out} =\langle\sigma_{\rm AGN}\rangle - <\langle\sigma_{\rm nAGN}>\rangle$.}

We find that  [O\,{\sc iii}] shows systematically higher velocity dispersion than [O\,{\sc i}].
%(Fig.~\ref{fig:sigo3o1}). 
This suggests that [O\,{\sc i}] and [O\,{\sc iii}] trace not only distinct gas phases, but also distinct phases of the outflow, and the relationship between [O\,{\sc iii}]] and [O\,{\sc i}] velocities is qualitatively consistent with [O\,{\sc iii}] tracing lower-density gas than [O\,{\sc i}], as expected -- the critical densities to produce the [O\,{\sc iii}]5007 and [O\,{\sc i}]6300 lines are $7\times10^{5}$ and  $2\times10^{6}$ cm$^{-3}$, respectively \citep{osterbrock06}.
%***Nadia: [some reference -- what are the typical densities of the [OI]-emitting gas?..]
%critical density of [OIII]5007= 7e5 /cm3.
%critical density of [OI]6300= 2e6 /cm3.

If the outflows result in shocks propagating from one phase to another, the densities and velocities of the different phases are related by $n_{\rm [OIII]} (v_{\rm out, [OIII]})^2 = n_{\rm [OI]} (v_{\rm out, [OI]})^2$. This implies that density of the [O\,{\sc i}] clouds is a factor 1.7 higher than that of the [O\,{\sc iii}] clouds. This result is consistent with theoretical predictions based on multi-component photoionization models of the NLR \citep{komossa97}. Assuming that the density of the clouds that produce the [O\,{\sc iii}] emission is 500\,cm$^{-3}$ -- a typical value of the electron density measured for AGN based on the [S\,{\sc ii}] emission lines \citep{dors14} -- we obtain $n_{\rm [OI]}\approx 850$\,cm$^{-3}$.  
%However, recent results suggest that the [S\,{\sc ii}] based measurements seem to underestimate the outflowing gas density by two orders of magnitude \citep{baron19}. Using the value derived for the electron density (10$^{4.5}$ cm$^{-3}$) by these authors as a proxy to $n_{\rm [OIII]}$, we estimate $n_{\rm [OI]}\approx5.4\times10^5$\,cm$^{-3}$. 
This value is smaller than the critical density for collisional de-excitation of the H$_2$ S(3) level of $\sim$10$^4$ cm$^{-3}$ \citep{roussel07}, and thus is consistent with our interpretation that the H$_2$ emission excess is likely produced by the same clouds that produce the  [O\,{\sc i}]$\lambda6300$. However, the nature of the outflows may be much more complex than our simple approach, as we are not able to properly constrain the geometries and gas densities of the outflows from single aperture spectra. Although the adopted value of $n_e$ is consistent with those derived for spatially resolved outflows using the [S\,{\sc ii}] emission lines \citep{couto16,lena15,lena16,soto19}, recent results suggest that the [S\,{\sc ii}]-based densities of ionized outflows can be underestimated by up to two orders of magnitude \citep{baron19b}.

%If the H$_2$ emission originates from a gas with density close to the critical density for collisional de-excitation of the S(3) level of $\sim$10$^4$ cm$^{-3}$ \citep{roussel07}, we estimate the warm gas outflow velocity of only $\sim20$\,km\,s$^{-1}$. This small velocity could explain the fact that hot molecular outflows from nearby AGN are still scarce. 

%However, recent results suggest that the [S\,{\sc ii}] based measurements seem to underestimate the outflowing gas density by two orders of magnitude \citep{baron19}. Using the value derived for the electron density (10$^{4.5}$ cm$^{-3}$) by these authors as a proxy to $n_{\rm [OIII]}$, we estimate $n_{\rm [OI]}\approx5.4\times10^5$\,cm$^{-3}$ and the expected velocity of the warm molecular phase would be $\sim180$\,km\,s$^{-1}$. For the extreme cases of our sample ($\sigma\approx250\,$km\,s$^{-1}$), the outflow velocity of the warm molecular gas would be as high as 500\,km\,s$^{-1}$, which is larger that that observed outflows velocities in hot H$_2$ \citep{eso,emonts17,may18}. 

\subsection{Energetics of the molecular outflows}

Our results indicate that shocks appear to play an important role in the production of both H$_2$ and [O\,{\sc i}]$\lambda6300$ emission. Theoretical models show that H$_2$ emission can be produced by shocks with velocities from 30 to 150 km\,s$^{-1}$ \citep{hollenbach89}, while the  shocks with velocities in the range 100--300 km\,s$^{-1}$ produce [O\,{\sc i}] emission \citep{ho14}. These values are smaller than the wind velocities (480--1500 km\,s$^{-1}$) derived from  hydrodynamical simulations of wind production from low-luminosity AGNs in sub-parsec scales \citep{almeida19}. Considering that the particles launched from the accretion disk are expected to decelerate due to the gravitational interaction at larger distances from the AGN, they can be responsible to produce the shocks needed to produce the H$_2$ and [O\,{\sc i}] emission.

Assuming a bi-conical geometry for the  molecular outflow, we can estimate the mass outflow rate through a circular cross-section with radius $r$ as
$\dot{M}_{\rm H2}=4\pi\,m_p\,f\,n_{H2}\,v_{H2}\,r^2$,
where $m_p$ is the proton mass, $f$ is the filling factor, $n_{H2}$ is the H$_2$ number density and $v_{H2}$ is the velocity of the H$_2$ outflow. Assuming a typical value for  bicone opening angle of $45^\circ$ \citep{ms11} to calculate $r$, $n_{H2}=850$ cm$^{-3}$ (as estimated for the [O\,{\sc i}] clouds), $f=0.01$ \citep[a typical value estimated for Sy galaxies, ][]{sb10,sm14}  and $v_{H2}\approx77$\,km\,s$^{-1}$ (as estimated for the [O\,{\sc i}]), the outflow rate at 200\,pc from the nucleus is $\dot{M}_{\rm H2}\approx1.6$~M$_\odot$ yr$^{-1}$.  We calculate the outflow rate at this distance because it corresponds to the peak of the location of the outflows in a sample of $\sim$4\,000 type 2 AGN \citep{baron19a}.  The derived mass outflow rate is consistent with those obtained from spatially resolved observations of outflows in nearby AGN \citep[e.g.,][]{diniz19,shimizu19}.

The kinetic power of the outflow is $\dot{E}_{\rm out}=\frac{1}{2}\dot{M}_{\rm H2}v_{\rm H2}^2\approx2.8\times10^{39}\,{\rm erg\,s^{-1}}$. The median [O\,{\sc iii}]$\lambda$5007 luminosity of the AGN in our sample is $\langle  L_{\rm [OIII]}\rangle=5.7\times10^{40}\,{\rm erg\,s^{-1}}$. Using a bolometric correction to the [O\,{\sc iii}]5007 luminosity of a factor of 3500 \citep{heckman04}, the corresponding bolometric luminosity is $L_{\rm bol}\approx2\times10^{44}\,{\rm erg\,s^{-1}}$. Thus, the kinetic power of the outflow is negligible compared to the AGN bolometric luminosity, meaning that there is no important feedback effect on the host galaxies.

The estimated velocities of warm molecular, neutral and ionized gas are $\lesssim100$~km\,s$^{-1}$, being smaller than the escape velocities of the galaxies of our sample. This implies in a ``maintenance mode" feedback, in which the gas is outflowing from the nucleus, but is re-distributed within the galaxies remaining available for further star formation. This result is similar to that found in other low-luminosity AGN \citep[e.g.][]{diniz19} and ULIRGs \citep[e.g.][]{emonts17}. Indeed, the power of the AGN outflows is strongly correlated with AGN luminosity  \citep{fiore17}.  In powerful AGN, such as luminous quasars, the velocity ($\gtrsim1000$ km\,s$^{-1}$) and kinetic power ($\times\,{\rm few}\, L_{\rm bol}$) of the outflow are much higher and then the  ``blow-out'' mode feedback takes place \citep{dimatteo05,hopkins05,sb10,hopkins12,zakamska16b,shimizu19}, in which the gas is expeled out of the galaxy.

\section{Conclusions and Implications}

In this work we match the sample of galaxies with mid-IR spectra available in the  {\it Spitzer} Telescope archive, compiled by \citet{lambrides},  with optical spectroscopic observations from the SDSS archive. From the 2,015 galaxies with mid-IR spectra, we find that 309 have SDSS spectra. This sample is used to investigate the origin of the excess of molecular hydrogen emission observed the mid-IR in nearby AGN host galaxies. By comparing mid-IR emission-line ratios with stellar populations properties, optical emission-line ratios and gas kinematics, we conclude that shocks play a major role in the production of the H$_2$ emission. These shocks are mainly due to AGN  driven winds.

We find strong correlations between H$_2$ fluxes and excitation temperatures and [O\,{\sc i}] fluxes and kinematics. Although similar relationships are also apparent between H$_2$ and [O\,{\sc iii}], these correlations are weaker. We interpret these relationships as evidence of AGN molecular outflows which we are indirectly uncovering using the [O\,{\sc i}] emission, which is a reliable tracer of shocks in neutral material. 

%Here, we find an indirect evidence of AGN molecular outflows. Such outflows are seems to play an important role in shaping galaxies, as they are claimed to  halt the star formation \citep{feruglio10,feruglio15,veilleux13,gonzalez-alfonso17} and thus affect the evolution of galaxies. The detection and characterization of such outflows in nearby galaxies is fundamental to properly constrain models of galaxy evolution and the role of AGN feedback \citep{cattaneo09,alexander12,fabian12,harrison17} and  verify if molecular and ionized gas outflows follow similar reltations with the AGN bolometric luminosity \citep{fiore17}. 

%However, (hot) molecular gas outflows are find to be scarce in nearby AGN, as revealed by spatially resolved near-IR observations  \citep[e.g][]{davies14,durre19,schonell19}. 
% Recent studies based on Atacama Large Millimeter/submillimiter Array (ALMA) observations of nearby galaxies reveal (cold) molecular gas outflows in some cases \citep{combes13,gb14,slater19,rama19}. So far, no systematic studies have been conducted to look for molecular gas outflows in nearby galaxies. The results presented here can be used to select galaxies to be likely hosts of molecular gas outflows, leading to important constrains of the AGN feedback role in  evolution of galaxies.

We find that objects with the strongest [O\,{\sc i}]$\lambda6300$/H$\alpha$ and highest velocity dispersions in [O\,{\sc i}] are the most likely hosts of molecular outflows. In order to confirm our hypothesis, these objects should be observed directly to get spatially resolved kinematics of the near-infrared ro-vibrational H$_2$ lines (which can be done using 10 m class ground based telescopes), mid-infrared rotational lines (can be done with JWST) and other molecular lines with ALMA to trace all different components of the outflow (hot, warm, cold).

\section*{Acknowledgements}
We acknowledge the referee for relevant suggestions that have
improved the paper. We thank Dr. S. B. Rembold for help with the CasJobs platform. This study was financed in part by Conselho Nacional de Desenvolvimento Cient\'ifico e Tecnol\'ogico (202582/2018-3, 304927/2017-1 and 400352/2016-8) and Funda\c c\~ao de Amparo \`a pesquisa do Estado do RS (17/2551-0001144-9 and 16/2551-0000251-7).

Funding for SDSS-III has been provided by the Alfred P. Sloan Foundation, the Participating Institutions, the National Science Foundation, and the U.S. Department of Energy Office of Science. The SDSS-III web site is http://www.sdss3.org/.

SDSS-III is managed by the Astrophysical Research Consortium for the Participating Institutions of the SDSS-III Collaboration including the University of Arizona, the Brazilian Participation Group, Brookhaven National Laboratory, Carnegie Mellon University, University of Florida, the French Participation Group, the German Participation Group, Harvard University, the Instituto de Astrofisica de Canarias, the Michigan State/Notre Dame/JINA Participation Group, Johns Hopkins University, Lawrence Berkeley National Laboratory, Max Planck Institute for Astrophysics, Max Planck Institute for Extraterrestrial Physics, New Mexico State University, New York University, Ohio State University, Pennsylvania State University, University of Portsmouth, Princeton University, the Spanish Participation Group, University of Tokyo, University of Utah, Vanderbilt University, University of Virginia, University of Washington, and Yale University.

%%%%%%%%%%%%%%%%%%%%%%%%%%%%%%%%%%%%%%%%%%%%%%%%%%

%%%%%%%%%%%%%%%%%%%% REFERENCES %%%%%%%%%%%%%%%%%%
% The best way to enter references is to use BibTeX:

%\bibliographystyle{mnras}
%\bibliography{example} % if your bibtex file is called example.bib

% Alternatively you could enter them by hand, like this:
% This method is tedious and prone to error if you have lots of references

%%%%%%%%%%%%%%%%%%%%%%%%%%%%%%%%%%%%%%%%%%%%%%%%%%

%%%%%%%%%%%%%%%%% APPENDICES %%%%%%%%%%%%%%%%%%%%%

\appendix

%\section{Some extra material}

%If you want to present additional material which would interrupt the flow of the main paper,
%it can be placed in an Appendix which appears after the list of references.

%%%%%%%%%%%%%%%%%%%%%%%%%%%%%%%%%%%%%%%%%%%%%%%%%%

% Don't change these lines
\bsp	% typesetting comment
\label{lastpage}
\end{document}